\documentclass[12pt]{article}

\usepackage{pslatex}
\usepackage{float}
\usepackage{amsmath}
\usepackage{color}
\usepackage{graphicx}
\usepackage{amssymb}
\usepackage{subfigure}
\providecommand{\tabularnewline}{\\}

\makeatother
\begin{document}

\begin{titlepage}
\thispagestyle{empty}
\begin{flushright}
{\bf  
IFJPAN-V-05-01\\   CERN-PH-TH/2005-091 \\
} 
\end{flushright}
\begin{center}
{
     \Large\bf
PHOTOS Monte Carlo: \; a precision tool  \\
 for QED corrections in $Z$ and $W$ decays}
\end{center}


\vspace*{2mm} 

\begin{center}
 
{\bf Piotr Golonka}\; and \; {\bf Zbigniew Was}
 \vspace{5mm}

{\em CERN, 1211 Geneva 23, Switzerland} \\ and \\
{\em Institute of Nuclear Physics, P.A.S., ul. Radzikowskiego 152, 31-342 Krak\'ow, Poland}

\end{center}
\vfill

\begin{abstract}
We present a discussion of the precision for the PHOTOS Monte Carlo algorithm,
with improved implementation of QED interference and multiple-photon 
radiation. 
The main application of PHOTOS is the generation of QED radiative corrections
in decays of any resonances, simulated by a ``host'' Monte Carlo generator.
By careful comparisons automated with the help of the MC-TESTER tool
specially tailored for that purpose, we found that the precision of
the current version of PHOTOS is of 0.1\% in the case of $Z$ and $W$ decays. 
In the general case, the precision of PHOTOS was also improved, 
but this will not be quantified here.
\end{abstract}

\vspace*{1mm}

\begin{center}
 {\em To be submitted to EPJC }
\end{center}

\vspace*{1mm}
\vfill
\begin{flushleft}
{\bf IFJPAN-V-05-01 \\
  CERN-PH-TH/2005-091 }
\end{flushleft}

\vspace*{1mm}
\bigskip
\footnoterule
\noindent
{\footnotesize \noindent
Supported in part  by the EU grant MTKD-CT-2004-510126, 
in partnership with the CERN Physics Department,
and the Polish State Committee for Scientific Research 
(KBN) grant 2 P03B 091 27 for years 2004-2006.
}
\end{titlepage}

\normalsize

\section{Introduction}

In the analysis of data from high-energy physics experiments, one tries to 
resolve the ``$experiment = theory$'' equation.  This no trivial task
requires that a lot of different effects be considered simultaneously.
From the experimental side, these are mainly detector acceptance and cuts,
which are dictated by the construction and physical properties of the detector: 
the shapes of distributions may be distorted by, say, misidentification 
and residual background contamination; these effects need to be discriminated 
in an appropriate and well-controlled way.
From the theoretical side, {\it all} effects of known physics have to be included in 
predictions as well. Only then can experimental data and theoretical predictions be 
confronted to determine numerical values of some coupling constants or 
effects of new physics (to be discovered). 

A well-defined class of theoretical effects consists of QED radiative 
corrections.
PHOTOS is a universal Monte Carlo algorithm  that simulates the effects of QED 
radiative corrections in decays of particles and resonances.
It is a project with a rather long history: the first version was
released in 1991 \cite{Barberio:1990ms}, followed by version 2.0 
\cite{Barberio:1994qi} in 1994 (double emission, threshold terms for fermions). 
The package is in wide use \cite{Dobbs:2004qw};
recently it was applied as a precision simulation tool for 
$W$ mass measurement at the Tevatron \cite{Abazov:2003sv}
and LEP  \cite{Abbiendi:2003jh,Abdallah:2003xn},
and for CKM  matrix measurements 
in decays of $K$ and $B$ resonances 
(NA48 \cite{Lai:2004bt}, KTeV\cite{Alexopoulos:2004up} , Belle \cite{Limosani:2005pi},
BaBar \cite{Aubert:2004te} and in Fermilab \cite{Link:2004vk}). 

Throughout the years the core algorithm for the generation of $O(\alpha)$ corrections
did not change much; however, its precision, applicability to various 
processes, and numerical stability improved significantly. 
Increased interest in the algorithm expressed by experimental
collaborations (including future LHC experiments) was a motivation to
perform a more detailed study of the potential and precision of the PHOTOS
algorithm; also new functionalities, such as multiple photon radiation
and better interference corrections, were recently introduced. The main purpose
of this paper is however not to document the new features of PHOTOS,
as this will be done elsewhere \cite{PGZWinprep,PhDGolonka},
but rather to present the results of precision tests of the algorithm.

Our paper is organized as follows: 
in section \ref{sec:Physical-problem} we present the physical problem
and further motivation for our studies. In section \ref{sec:PHOTOS-features}
we recapitulate the history and properties of the PHOTOS algorithm. In
section \ref{sec:Test-definition} we present the testing procedure
and necessary tools we employed to obtain our main results, which are presented
in section \ref{sec:Results}. Finally, section \ref{sec:Summary-and-Outlook}
summarizes the paper and gives an outlook to further developments.
In Appendix \ref{sec:new-photos-features} we describe the most important features 
of the new version of the PHOTOS code.

Some results of the tests and improvements of PHOTOS, relevant 
to $\tau$ physics, have already been discussed in \cite{Was:2004dg}.

%
\section{\label{sec:Physical-problem}Physical problem and developed solution}
%

Although QED bremsstrahlung in particle decays is one of the most elementary effects
in quantum mechanics, it is not usually considered explicitly; to simplify,
calculations are performed for inclusive quantities.
In fact, it is only the case of a few specific decay channels, where
exact fixed-order (e.g. $O(\alpha^{2})$) fully-differential formulae,
with spin amplitudes or matrix elements squared, are available in analytical, 
semi-analytical or Monte-Carlo form. Nonetheless, in the analysis of the experimental data,
radiative corrections are usually treated together with the detector effects 
(e.g. conversion, detector efficiency) to form the ``QED-subtracted'' data. 
The control of the uncertainties of such data becomes a weak point:
theoretical uncertainties appear on both sides of the
``$theory=experiment$'' equation \cite{Was:1994kg}.
This problem does not seem to be so evident for ``discovery''
experiments or measurements performed on small statistical samples:
usually the effects of radiative corrections do not exceed a few
per cent. An exception is radiative corrections for background processes,
such as $gg \to t \bar{t}; \; t \to l \nu_l b$, where QED bremsstrahlung is 
an essential element for this channel background contribution to Higgs boson searches
at the LHC in the $H \to \gamma \gamma$ channel \cite{Richter-Was:1994ep,Richter-Was:1993ta}.
Conversely, for high-precision measurements (as those
performed nowadays), good control of the radiative corrections becomes
vital not only for the assessment of the overall experimental error of
the respective cross-sections or branching ratios, but
also for the shapes of the distributions.

The effects of radiative corrections gradually became an important
topic in the context of such measurements as high-precision measurements 
of $W$-boson properties (see, for instance, the combined results of Tevatron runs 
\cite{Abazov:2003sv}), or $B$, $D$, $K$ meson decays for measurements
of CKM-matrix coefficients in $B$-physics \cite{LaJolla2005}.
With increasing statistics of available experimental data, QED radiative
corrections have become a significant element in the systematic error of
the measured quantities.

Strict, systematic calculations performed order-by-order in the perturbation
theory are usually not the most efficient way of including the effects
of bremsstrahlung. To improve the convergence of the perturbative expansion,
the most popular method in QED is exponentiation, a rigorous scheme of 
reshuffling the dominant terms between orders of expansion. 
This method is useful for the construction of Monte Carlo algorithms as well 
\cite{kkcpc:1999,koralz4:1994}.
In the leading-log approximation, partially inclusive formulae 
exhibit factorization properties of QED, see e.g. \cite{Eberhard:1989ve}. 
A matrix element formula for particle decay accompanied by bremsstrahlung 
photon emission may be factorized to Born-level terms times 
the bremsstrahlung factor%
\footnote{For some cases this factorization property is also present for 
non-approximated formulae, e.g. for the $O(\alpha)$ ME formula for $Z$-boson 
decays as implemented in MUSTRAAL \cite{Berends:1982ie,Berends:1983mi}
at fully differential level. In \cite{Barberio:1990ms} it was shown, 
contrary to the previous expectations, that this factorization has a natural 
geometrical interpretation as well.}.

Similarly, a fully differential formula for Lorentz-invariant phase space
for particle decay accompanied by a number of photons may be expressed
in a way that exhibits factorization properties. It is used, for example, 
in the construction of the TAUOLA algorithm for 
$l \nu_\tau \bar{\nu}_l (\gamma)$ \cite{Jezabek:1991qp}. 
The important property of this parametrization is the full coverage
of the phase space and exact (i.e. free of approximations) treatment. 
The price to pay is that the variables describing each added photon
are defined in individual rest frames, separated by boosts, 
making the question of the choice of gauge fixing a very subtle point of the 
theoretical bases of the algorithm \cite{Was:2004ig}.

It is nonetheless straightforward to take advantage
of these factorization properties and approximate the fully differential
formula for the cross-section in any particle-decay process accompanied
by a bremsstrahlung photon, by a product of a cross-section formula that
does not contain QED radiative corrections times a bremsstrahlung
factor. The bremsstrahlung factor depends only on the four-momenta 
of those particles taking part in the decay, and not on the actual underlying process! 
This approximation, which takes into account both real and virtual
corrections, converges to an exact expression in the soft-photon region
of phase space; this fact is exploited in the construction of PHOTOS.
It exposes well-known infrared and collinear singularities.
In the final formula, the infrared divergences that originate from 
expressions describing the emission of real and virtual photons may 
be regularized and cancelled out order-by-order. 
To realize this operation, a technical parameter giving the minimum photon
energy is defined; then, integration over the directions of photons
with energies lower than the cut-off is performed; this results in
the final formula being free from infrared regulator. 
Only photons of energies higher than the cut-off
are explicitly generated%
\footnote{
From a practical point of view this poses no problem: the energetic
resolution of calorimeters used in the experiments is also limited.}.
The PHOTOS algorithm properly treats the collinear region of the phase space
as well: the singularities are regulated simply by the masses of the charged 
particles.  

As a result of the studies of the second-order matrix element,
in particular for $Z\rightarrow\mu^{+}\mu^{-}\gamma\gamma$ and 
$gg\rightarrow t\bar{t}\gamma\gamma$ (performed already in 1994 
\cite{Richter-Was:1994ep,Richter-Was:1993ta}), the iterative properties
of the bremsstrahlung-adding formula have been identified and a universal
photon-emission kernel has been isolated and double-photon
emission implemented in PHOTOS. 
However, at that time, the question of precision was left out of
the discussion. The main application was an estimate of the background to
the Higgs boson searches in the $H \rightarrow \gamma \gamma$ channel.
Only double hard-photon bremsstrahlung was of interest,
and a precision of about 10\% on the cross-section was sufficient. 

In the iterative algorithm for (fixed-number) multiple-photon emission in PHOTOS,
the probability of photon generation is based  (at crude level) 
on a binomial distribution, i.e. the probability for generating 
and not generating a photon is calculated in each iteration.
Encouraged by the precision of PHOTOS with a photon multiplicity higher than 2
(subsequent terms in the expansion improved the agreement 
with exact calculations), we decided to replace the binomial distribution with 
a Poissonian distribution for the number of photons to be generated.
In this new Poissonian mode, the algorithm is no longer limited to a fixed maximal 
number of photons: the actual number of photons is also generated 
(at this ``crude-generation'' level).
The modification set the algorithm free from the negative-probability problem as well.
At a crude level, the total probability for single-photon emission in the iterated kernel 
does not need to be smaller than 1, because it is multiplied 
by the exponent of an equally large, but negative, number\footnote{This 
is trivial: if $p$ is the probability in binomial distribution,
it must be smaller than 1; however it does not need to have in Poissonian
distribution given by $ P_n = \frac{1}{n!} {e^{-p} p^{n} } $ 
}. 
Because the crude distribution resembles an expansion of an exponent
we call this version of multiple-photon algorithm an ``exponentiated mode'' 
of PHOTOS.
Our approach is also close to the language of exponentiation as known
in QED since Yennie-Frautchi-Suura times \cite{yfs:1961}.
The exponentiated mode has proved to be more stable, allowing us to 
significantly lower the value of the infrared cut-off on the photon energy as well. 

A flexible organisation of iteration and phase-space variables in the PHOTOS
algorithm not only allows a full phase space coverage, but also seems
to introduce some higher-order effects, as will be discussed 
later in the paper (subsection \ref{sub:Iterating-emission-kernels}). 
This was achieved by careful studies and comparisons with matrix-element 
calculations \cite{Richter-Was:1994ep,Richter-Was:1993ta,Was:2004ig},
without need of any kind of phase-space ordering.

\section{\label{sec:PHOTOS-features} Evolution of the main features of the PHOTOS algorithm}

PHOTOS is an ``after-burner'' algorithm, which adds bremsstrahlung photons to 
already existing events, filled in by a host generator (which does not take
into account the effects of QED radiative corrections) and transmitted by means
of a standard HEPEVT event record (only the information about four-vectors
of particles taking part in the process, and the topology of the process
are needed). PHOTOS adds (with a certain probability) final-state 
QED radiation in ``any'' decay of particle or resonance,
independent of the physics process that was generated.
As the result of its execution, bremsstrahlung photons are added in
a fraction of the events in the HEPEVT event record, taking
into account event topology and momentum conservation.

The single-photon version of the PHOTOS algorithm originates from the MUSTRAAL
Monte Carlo \cite{Berends:1982ie,Berends:1983mi}, 
in its part for the $Z\rightarrow\mu^{+}\mu^{-}(\gamma)$ process.
In PHOTOS the full $O(\alpha)$ matrix-element algorithm was 
simplified so as to isolate the universal kernel
responsible for photon emission. The original algorithm, although maintaining
full $O(\alpha)$ precision and exposing factorization properties,
was however dependent on the underlying process%
\footnote{It remains implemented in the first published version of KORALZ
3.8 \cite{Jadach:1991ws}, with multiple initial-state bremsstrahlung and single 
final-state radiation.}.
A downgrade was therefore required to make the algorithm 
process-independent: the interference terms were removed from the formula,
and later restored in an approximated way for a limited set of processes (decays
of neutral particles into two charged particles of the same mass) by
introducing a Monte Carlo interference weight. 
The first version of PHOTOS: universal Monte Carlo for QED radiative corrections 
in particle decays was released in 1991 \cite{Barberio:1990ms}. 
The algorithm operated on the four-vectors stored, in the HEPEVT event record
\cite{Hernandez:1990yc}, by a host generator.

In 1994, version 2.0 of PHOTOS was released
\cite{Barberio:1994qi}. The most important improvement was an
implementation of the double-photon emission obtained by iteration of
the bremsstrahlung kernel.  This version became widely used in the HEP
experimental era \cite{Dobbs:2004qw}. 
In 2000, PHOTOS was integrated to the TAUOLA-PHOTOS-F environment
to ease code maintaining; the update of PHOTOS was released
as part of the documentation of that package \cite{Golonka:2000iu}.
In 2003, driven by the needs of experimental collaborations, a dedicated
Monte Carlo weight for $W\rightarrow l\nu(\gamma)$ process was
developed \cite{Nanava:2003cg}. 
This new option, implemented in PHOTOS version 2.07, was documented 
in \cite{Golonka:2003xt}. 
One of the changes performed at that time, 
trivial from the physical point of view, but turned out 
recently to be vital, was the change of variables in the calculation of 
interference weight in PHOTOS: from angles to four-vectors.

Problems with numerical stability may occur when the four-vectors passed in 
the event record have insufficient precision.
PHOTOS is very sensitive to rounding errors and momentum conservation
in the event record, especially when the multiple-photon-emission mode of operation is used.
At each iteration of the photon-emission kernel numerical rounding errors are
accumulated. The value of the infrared cut-off parameter was raised to
protect the algorithm from being stopped because of numerical instability.
It is not the desired method to solve the numerical problems of course. 
Special kinematic-correcting subroutines have been made available in 
PHOTOS since version 2.07, but they are not active in the default setting 
of input parameters.

In July 2004, we started a systematic study of the PHOTOS algorithm,
focusing on its precision and possible extension to the multiple-photon
emission. It turned out that the algorithm stability suffered from
a bug; this had no impact on the results, but prevented the
photon emission kernel to be iterated more that twice. Once the bug
had been identified and corrected, the algorithm was extended
to triple- and quartic-photon emission, then an exponentiated version
of an iteration routine was implemented.

In January 2005, driven by needs of $B$-physics experiments,
a new, universal interference weight was implemented, allowing for
the calculation of an interference for ``any'' process (see also subsection
\ref{sub:photos-B}). Although the formula for this interference weight
is given already in eq. (17) of \cite{Barberio:1994qi},
its integration with the PHOTOS code was not trivial until recently; it was
modification made in 2003 that made the implementation possible
for every decay.

From the point of view of the precision of the obtained results,
the PHOTOS algorithm may work in three regimes: (1) as a ``crude-level'' tool
for bremsstrahlung generation in decays of any particle or resonance,
(2) as a precision tool for dedicated decay channels (at present $Z$ and $W$ decays), 
(3) as a precision tool incorporating matrix-element calculations
(not exploited so far). 

At a ``crude level'',  only the leading-log soft-photon 
parts of the matrix element are included in the PHOTOS algorithm%
\footnote{ It can be estimated
that the physical uncertainty of the PHOTOS
algorithm, with double emission and working on the decay of particle $P$ into a charged
particle $ch$ and the neutral system $Y_i$,
is then not smaller than $\frac{\alpha}{\pi}$ or
$(\frac{\alpha}{\pi}\log \frac{m_{P}^2}{m_{ch}^2})^3$, whichever is
bigger.}.  
PHOTOS is most often used as a general-purpose tool, thus working at the 
``crude level'' of precision. The precision of the results cannot be
guaranteed in this working regime; a typical example of use
could, however, be the generation of full bremsstrahlung 
phase-space coverage for acceptance studies only.
Appropriate comparisons with matrix-element calculations 
or experimental data would have to be performed to determine the actual precision.
For the particular decay channel, the PHOTOS algorithm could then be upgraded to regime 
(2) or (3), depending on the form of necessary compensating weight. 
At present, the algorithm  employs  a number of improvements 
(with the help of the correcting weight)  and takes into account 
such effects as the threshold terms for fermions and the impact of 
the spin of emitting particle. 
Other effects (e.g. the impact of spin of a decaying particle or the influence 
of spin on interference terms) are not at present taken into account.

The impact of the approximations and missing terms on the precision of PHOTOS,
when used in regime (2), that is for $W$ and $Z$ decays, 
is the main subject of this paper.

\section{\label{sec:Test-definition}Test definition}

Our approach to the study of precision of the PHOTOS algorithm
presented in this paper is based on numerical comparisons of 
results obtained from PHOTOS with respect to the results produced by other 
{\it reference} Monte Carlo generators.
The reference generators that we used employ formulae based on the full 
(i.e. non-approximated) fixed-order matrix-element calculations
with or without exponentiation;
their precision level is well established, both by theoretical
considerations and by comparisons with experimental data. 
The advantage of such an approach is obvious: it allows us
not only to check the theoretical precision of the 
approximation used, but also verifies that there 
are no accidental errors in the actual computing code.

This approach is probably more appealing to the practically-oriented user 
than to the theoretically-oriented one. We leave more profound theoretical
studies for the future, to be possibly performed in the context of feasibility
studies of applicability of methods used in PHOTOS to QCD.

We have tested PHOTOS against a limited number of event generators 
and for a limited number of processes. 
The choice of processes was dictated not only 
by physic interests, but also by the availability of Monte Carlo 
generators with well-controlled and established precision. 
The following processes were studied:
\begin{itemize}
\item $Z^{0}\rightarrow\mu^{-}\mu^{+}$: for this process the LEP era generators
 KORALZ \cite{koralz4:1994,koralz4:1999} and KKMC \cite{kkcpc:1999}
could be used. Note that these programs agreed well with practically 
all experimental data of the LEP measurements. The KKMC generator is based
on $O(\alpha^2)$ ME calculations with spin amplitudes technique
and exponentiation. It can also be used in $O(\alpha)$ ME
 exponentiated mode.
In the case of KORALZ,  generation at first-order matrix element and
no exponentiation were available as well.
\item $W^{+}\rightarrow\mu^{+}\nu_{\mu}(\gamma)$: for this process the WINHAC \cite{Placzek:2003zg}
generator is available for multiple photon generation in decay. It is based
on first-order matrix element and exponentiation. Comparison of PHOTOS with
a first-order matrix-element generator without exponentiation
can be found in ref.~\cite{Nanava:2003cg}; we will not repeat it here.
\item $H\rightarrow\mu^{+}\mu^{-}(\gamma)$: comparison of PHOTOS with
a first-order matrix-element generator without exponentiation 
can be found in ref.~\cite{Andonov:2002mx}; we will not repeat it here.
\item for leptonic $\tau$-decays, complete QED first-order 
generation with the TAUOLA \cite{Jezabek:1991qp} generator
was available for tests; TAUOLA was widely used and compared successfully with LEP 
and CLEO data \cite{Davier:2002mn,Anderson:1999ui}.
\end{itemize}
The event generators and physical initializations
mentioned above will be referred to as ``reference generators''.
They were used to produce reference results used in comparisons with PHOTOS.

The testbed constructed from the event generators discussed above had one very
important feature: it allowed a test of PHOTOS against several algorithms 
that differ in physical content and precision.
Tests started from an exact $O(\alpha)$ matrix element for single-photon emission
(non-exponentiated KORALZ and TAUOLA) compared with a single-photon version
of PHOTOS), and went on to multiple-photon generators (exponentiated
versions of KKMC, KORALZ and WINHAC) compared with triple-, quartic-
and exponentiated multiple-photon modes of PHOTOS.

To operate, PHOTOS needs events produced by a ``host'' generator as an input.
We exploited the possibility of deactivating the bremsstrahlung generation 
in reference generators, to turn them into QED Born-level ``host'' 
generators.

Let us now define the method for automated tests that were
used to obtain the results presented later in this paper. To facilitate
the systematic comparisons, we adopted a specially designed tool: 
MC-TESTER, described in detail in ref. \cite{Golonka:2002rz}.
The principle of an underlying test is to analyse the series of events 
produced by two distinct Monte Carlo event generators 
(or a sequence of several generators combined together) and 
to extract characteristic distributions in an automatic way.
In practice, we search for distinct decay channels of the particle 
of interest and store histograms with distributions of all 
invariant masses that can be formed from the four-momenta of 
its decay products.

For a selected decay process, in cases discussed here, $Z$, $W$ and $\tau$ decays,
the distributions are extracted from the event record in an automated way (thereby
limiting manpower for setting up the appropriate analysis code, and
also the risk of accidental errors) and stored in the output files.
Two output files (from distinct runs of event generators instrumented
with MC-TESTER) are then analysed, and the results are presented in
a form visualized  as a ``booklet'' of plots and tables.
The user is given a general information concerning two compared runs of
Monte Carlo generators, a list of the decay channels with their branching
fractions, and the maximum values (for each decay channel)
 of the Shape Difference Parameter (SDP)%
\footnote{ The Shape Difference Parameter defined in \cite{Golonka:2002rz}
quantifies the difference in shape of the histograms coming from 
two compared runs. The histograms contain distributions of all possible
invariant masses, which can be constructed from the momenta of the decay 
products of the particle under study; hence there is a set of histograms
for every decay channel. The SDP value is calculated for each histogrammed
mass: it quantifies the exclusive surface between the 
(normalized to unity) corresponding histograms obtained from the two runs.
The effects of statistical fluctuations are appropriately subtracted.
The maximum of SDP over all distributions for a given decay channel is taken.}. 
For each decay channel the plots of histogrammed values are then included:
each plot presents the two distributions from two distinct runs, and the
curve being a ratio of the normalized distributions; the value of the SDP
is also printed for each plot.

The testing approach implemented in MC-TESTER was already very useful in the 
case of validation of TAUOLA package; however, for the purpose discussed here
it required further adaptation. The problem was in consistent treatment of
the arbitrary number of final-state QED bremsstrahlung photons that might be 
present in the event, the ambiguity caused by infrared singularities 
of QED, which is handled differently by the various Monte Carlo programs.

Our aim was to develop a technique where comparisons would make physical sense,
would be automatic and independent of the way infrared singularities 
are regularized in particular generators. 
For our analysis, we defined zero-, one-, and two-photon topologies in
the following way: we called the event ``zero photon''
if there was no photon of energy (in a decaying particle rest-frame)
larger than the parameter $E_\mathrm{test}$.
The ``one-photon'' event had to have one (and only one) photon of 
energy larger than $E_\mathrm{test}$. 
If there were more than one such photons, we called it a ``two-photon'' event.
If there were more than two photons of energy larger than $E_\mathrm{test}$, 
we considered only the two most energetic ones, and treated
the remaining ones as if they had not passed the $E_\mathrm{test}$ threshold.
For all the photons that did not pass the $E_\mathrm{test}$ threshold we
performed an operation inspired by leading-log logic: we
added their four-momenta to the momenta of outgoing fermions of smaller
angular separation.

We defined two variants for this test definition:
\emph{test1} and \emph{test2}. The \emph{test2} was
exactly as explained above. In \emph{test1}, only one photon (the
most energetic one) could be accepted%
\footnote{ If needed, this scheme could be
generalized, for instance, to define \emph{test3}, where up to three 
photons of energies above $E_\mathrm{test}$ could be accepted, leading to 
a fourth distinct decay channel.}.
The free parameter of the test:
$E_\mathrm{test}$ was adjusted for each process so that the results had
physical sense.

Because of the space limitation we cannot present complete ``booklets'' 
with results of comparisons; for the purpose of this paper we decided
to extract the most vital information: the branching ratios for events 
with 0, 1 and 2 photons (0 and 1 if \emph{test1} was used) 
and the maximum value of SDP throughout all the plots in a channel. 
Complete results with summary tables and all analysis booklets are 
available on the web \cite{tauolaphotos}. 
The tests were based  on high-statistics runs ($10^{8}$ non-weighted
events) for all generators. 

Finally, let us stress that an essential preliminary step for the tests 
presented here was to assure the numerical stability of PHOTOS. 
As a consequence, we could determine that PHOTOS may be used for processes at very 
high energies (i.e. at the scale of LHC energies and even at astrophysics ones).
The infrared cut on the photon energy may be lowered to
$10^{-7}$ from the $10^{-2}$ value that had to be used before.

\section{\label{sec:Results} Numerical results}
The results presented here cover two issues:
precision of the predictions obtained for the single-photon emission kernel
and the convergence of the iterative solution.
At first, in subsection \ref{sub:Single-photon-emission-kernel-tests},
 we will discuss the comparisons of a single-photon emission kernel
implemented in PHOTOS with other genuine first-order generators.
In subsection \ref{sub:Iterating-emission-kernels},
we will discuss tests for multiple-photon generation.
Finally, in subsection \ref{sub:photos-high-energies}, we will discuss the applicability
of PHOTOS at very high energies.

As mentioned before,  we aim at comparing, in total,
{\it six} versions of distinct Monte Carlo programs with five distinct 
modes of executing PHOTOS. Let us specify them here (including acronyms
that will be used throughout the text):
\begin{itemize}
\item { \bf KORALZ O(1)}: KORALZ generator \cite{koralz4:1999} with $O(\alpha)$ ME for $Z^0 \rightarrow \mu^+ \mu^- (\gamma)$, no exponentiation, single-photon emission
\item { \bf KORALZ     }: KORALZ generator \cite{koralz4:1999} with $O(\alpha^2)$ ME for $Z^0 \rightarrow \mu^+ \mu^- (\gamma)$ and exponentiation (multiple-photon emission)
\item { \bf KKMC       }: KKMC   generator \cite{kkcpc:1999}   with $O(\alpha^2)$ ME for $Z^0 \rightarrow \mu^+ \mu^- (\gamma)$ and exponentiation at spin-amplitudes level
\item { \bf KKMC $ O(1)_\mathrm{EXP} $ }: KKMC generator \cite{kkcpc:1999} with $O(\alpha)$ ME for $Z^0 \rightarrow \mu^+ \mu^- (\gamma)$ and exponentiation
\item { \bf TAUOLA     }: TAUOLA generator \cite{Jezabek:1991qp} with $O(\alpha)$ ME for $\tau \rightarrow l \bar{\nu}_l \nu_\tau (\gamma)$ (single-photon emission)
\item { \bf WINHAC EXP }: WINHAC generator \cite{Placzek:2003zg} with full $O(\alpha)$ ME for $W \rightarrow l \bar{\nu}_l (\gamma)$ and exponentiation
\item { \bf PHOTOS O(1)}: PHOTOS, no iteration
\item { \bf PHOTOS O(2)}: PHOTOS algorithm iteration up to two times
\item { \bf PHOTOS O(3)}: PHOTOS algorithm iteration up to three times
\item { \bf PHOTOS O(4)}: PHOTOS algorithm iteration up to four times
\item { \bf PHOTOS EXP }: PHOTOS algorithm with exponentiation, multiple-photon generation
\end{itemize}

Before we proceed with a presentation of the results, let us, however,
briefly describe the approach taken for this presentation and their numerical quantification:
\begin{itemize}
\item As described in section \ref{sec:Test-definition}, we quantify 
the difference between results produced by two generators by 
calculating branching ratios for events with 0, 1 or 2 photons
with energy above the $E_\mathrm{test}$ threshold, and the maximum values of the SDP parameter
of all combinations of invariant masses in the specific channel 
(with 0, 1 or 2 photons) for a given comparison.
\item In discussion of the results we will also often use a single value of
the ``overall difference'', being either the maximum of differences 
in branching ratios
or the maximum of products of SDP and
corresponding branching ratios\footnote{
For instance, in case of the table in Fig.~\ref{fig:O1ResultsKoralPhotosO1} 
the difference in branching ratios is $0.82514-0.82362 = 0.00152$, 
the product $\mathrm{SDP} \times \mathrm{BR} = 0.0053*0.176 = 0.00093$, so that the first 
value is taken as the overall difference for that test: 0.152\%.}.
We then take the larger of the two.
\end{itemize}

Later in this section we present only the simplified  summary tables 
and plots where the differences in distributions 
(quantified by the SDP parameter) are most significant.
In the summary tables the first line (appearing in bold font) refers
to the results produced by the reference generator, the others
refer to results of runs of generators being tested.

\subsection{\label{sub:Single-photon-emission-kernel-tests}Single-photon emission kernel}

In this subsection we focus on the quality of the single-photon 
emission kernel in the PHOTOS algorithm. 
We perform a systematic comparison of PHOTOS
with generators that implement $O(\alpha)$
matrix element for $Z$, $W$ and $\tau$ decays.

The purpose of these tests was to assess the impact of the 
simplifications of the photon-emission kernel implemented in PHOTOS
on the precision of its predictions 
(in the PHOTOS kernel the non-leading $O(\alpha)$ terms are omitted).

At first, let us present the comparisons made in the $Z\rightarrow\mu\mu(\gamma)$ channel.
In fact, the original PHOTOS algorithm was created by gradual
simplification of the exact ME formula for this process, so the universal
emission kernel could have been identified. Therefore this
 process remains a benchmark for PHOTOS single-emission algorithm, 
its precision and design. 
The approximations are well understood and may in fact be restored 
for this particular case. 

\begin{figure*}
\caption{\label{fig:O1ResultsKoralPhotosO1} Predictions of KORALZ 
(with $O(\alpha)$ matrix-element and single-photon emission) are compared with predictions of PHOTOS 
(running in single-photon option) for the $Z^{0}\rightarrow\mu^{+}\mu^{-}(\gamma)$ 
channel, $E_\mathrm{test}=1.0$ GeV. 
The plot presents the distribution of invariant mass of $\mu^- \mu^+$ pair 
coming from KORALZ (in red, or darker-grey) and PHOTOS (in green, or lighter-grey); 
the black line is the ratio of the two normalized distributions.
The red and green lines are hard to separate - they practically overlap.}

\begin{center}
\begin{minipage}[c]{1.0\textwidth}%
\begin{center}~\\
\begin{tabular}{|c|c|c|c|c|}
\hline 
GENERATOR           & \multicolumn{2}{c|}{Branching ratio} & \multicolumn{2}{c|}{Max SDP} \tabularnewline
$n$ photons $\to$     &             0  & 1                   &   0    & 1                   \tabularnewline
\hline
\hline 
\textbf{KORALZ O(1)}& \textbf{0.82514}& \textbf{0.17486}     & \multicolumn{2}{c|}{}        \tabularnewline
\hline 
PHOTOS O(1)         &      0.82362    & 0.17638              &      0 & 0.0053              \tabularnewline
\hline
\end{tabular}\\
~\\
~\\
\includegraphics[width=0.60\textwidth]{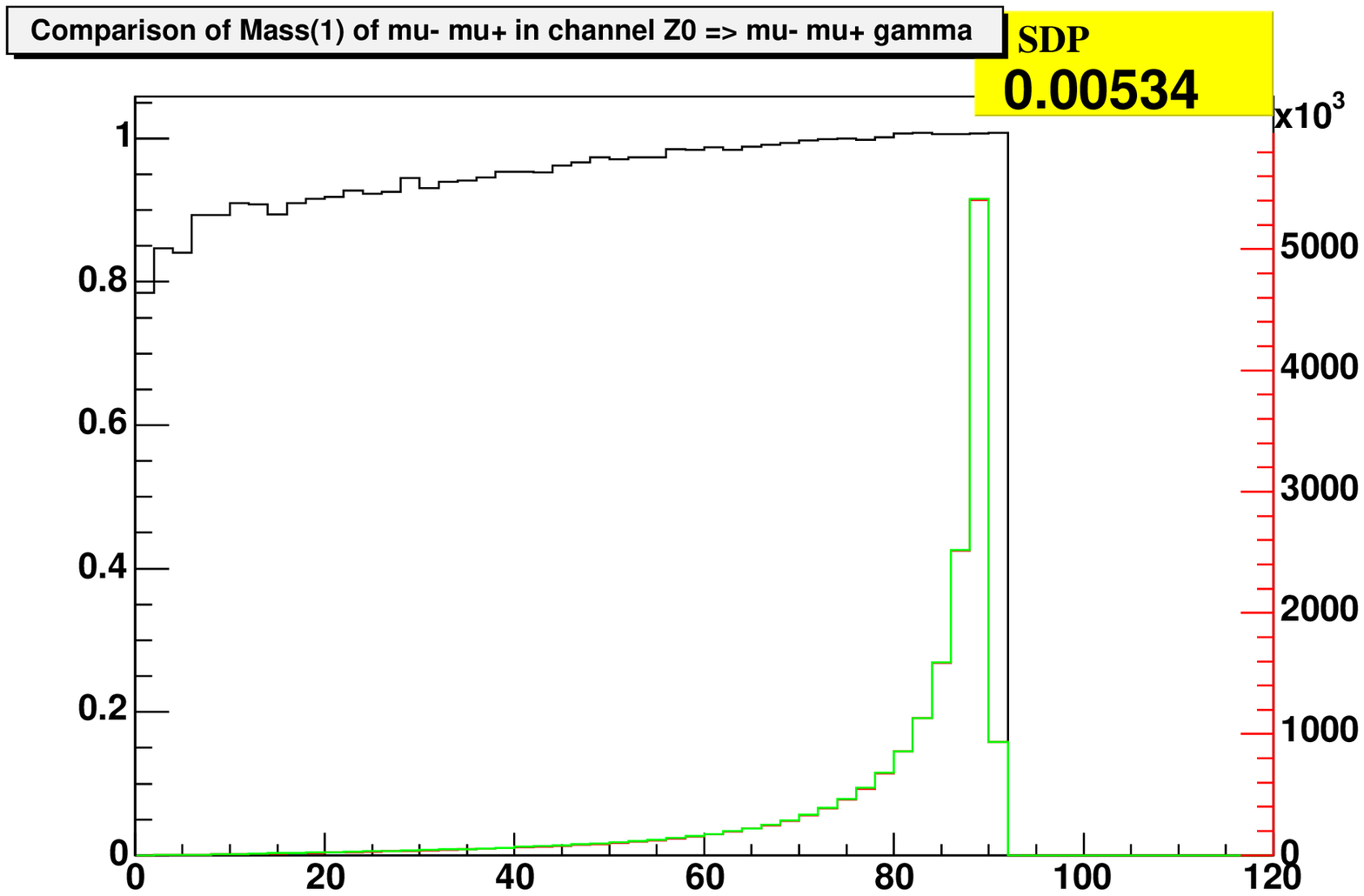}
\end{center}
\end{minipage}%
\end{center}
\end{figure*}

We compare PHOTOS with KORALZ Monte Carlo in single-photon
emission mode. This test, which was already performed a long time ago 
(when PHOTOS was being developed), has been reproduced now, and confirmed 
that no accidental error has occurred since then.
The results of this comparison are presented 
in Fig.~\ref{fig:O1ResultsKoralPhotosO1}. 
One may notice that the difference in branching ratios for channel
with and without photon are at the per mil level.  
The differences in distributions (the maximum value of SDP multiplied by
branching ratios)  indicate agreement better than per-mil: 
we have quantified the overall difference as 0.15\%.

The dominant contribution to the difference originates from a very sparsely
populated area of the phase space, where the invariant mass of the lepton system
is small, and very hard photons were emitted. 
The discrepancy between the distributions does not exceed 20\%, even
in this corner of the phase space, though.
It is visible in the slope of the black  curve denoting the ratio of the 
two distributions. 
In this region of the phase space the approximations 
(with respect to $O(\alpha)$) used in PHOTOS are 
the largest. The differences are too small to justify the implementation of 
a channel-dependent correction weight.

For the case of $W$-boson decays, we address the reader to
\cite{Nanava:2003cg}, where the comparison of PHOTOS and a $O(\alpha)$ ME
generator SANC~\cite{Andonov:2002jg}\footnote{SANC calculates complete one-loop amplitudes
for the decays of on-shell vector bosons: $W$, $Z$ or the Standard Model Higgs boson; 
an exact, single, real-photon emission matrix element is obtained that way.}
 for the $W\rightarrow \mu \bar{\nu}_mu (\gamma)$ process is given. 
This paper provides also the theoretical background for the $W$
interference weight implemented in 2003 in PHOTOS 2.07. 
Presented tests indicate very good agreement between PHOTOS and SANC:
up to a level of 1\% (statistical error) in the majority of the areas of the phase space, 
within 5\% for parts of distributions where collinear-induced logarithms are
absent and 10\% in the regions where only non-leading corrections
contribute to the matrix element.

The paper \cite{Andonov:2002mx} presents the comparison of various distributions
from PHOTOS and SANC~\cite{Andonov:2002jg} generators for Higgs boson decays:
an agreement at the level of 1\% (statistical error) was found all over the phase space. 

We consider the results of these tests, published in relatively recent papers 
\cite{Nanava:2003cg, Andonov:2002mx}, as being relevant and complementary to the tests 
presented here. Therefore we do  not repeat the comparison of 
WINHAC generator \cite{Placzek:2003zg} and PHOTOS in single-photon 
emission mode. 

\begin{figure*}

\caption{\label{fig:O1ResultsTAUOLAvsPHOTOSmu}Predictions for 
exact $O(\alpha)$ in TAUOLA  are compared with single-photon emission of PHOTOS in 
the $\tau^- \rightarrow \mu^-  \bar{\nu}_\mu \nu_\tau (\gamma)$ channel, 
$E_\mathrm{test}=0.05$ GeV.
The plot presents the distribution
of invariant mass of the $ \bar{\nu}_\mu \nu_\tau \gamma$ system coming from 
TAUOLA (in red, or darker-grey) and  PHOTOS (in green, or lighter-grey); 
the black curve is the ratio of the two normalized distributions.}

\begin{center}\begin{minipage}[c]{1.0\textwidth}%
\begin{center}\begin{tabular}{|c|c|c|c|c|}
\hline 
GENERATOR& \multicolumn{2}{c|}{Branching ratio}& \multicolumn{2}{c|}{Max SDP} \tabularnewline
$n$ photons $\to$ &0&1&0&1\tabularnewline
\hline
\hline 
\textbf{TAUOLA}&\textbf{0.98916}&\textbf{0.01084}&\multicolumn{2}{c|}{}\tabularnewline
\hline 
PHOTOS O(1)&0.98927&0.01073&0&0.0044\tabularnewline
\hline
\end{tabular}\\
~\\
~\end{center}

\begin{center}\includegraphics[%
  width=0.60\textwidth,
  keepaspectratio]{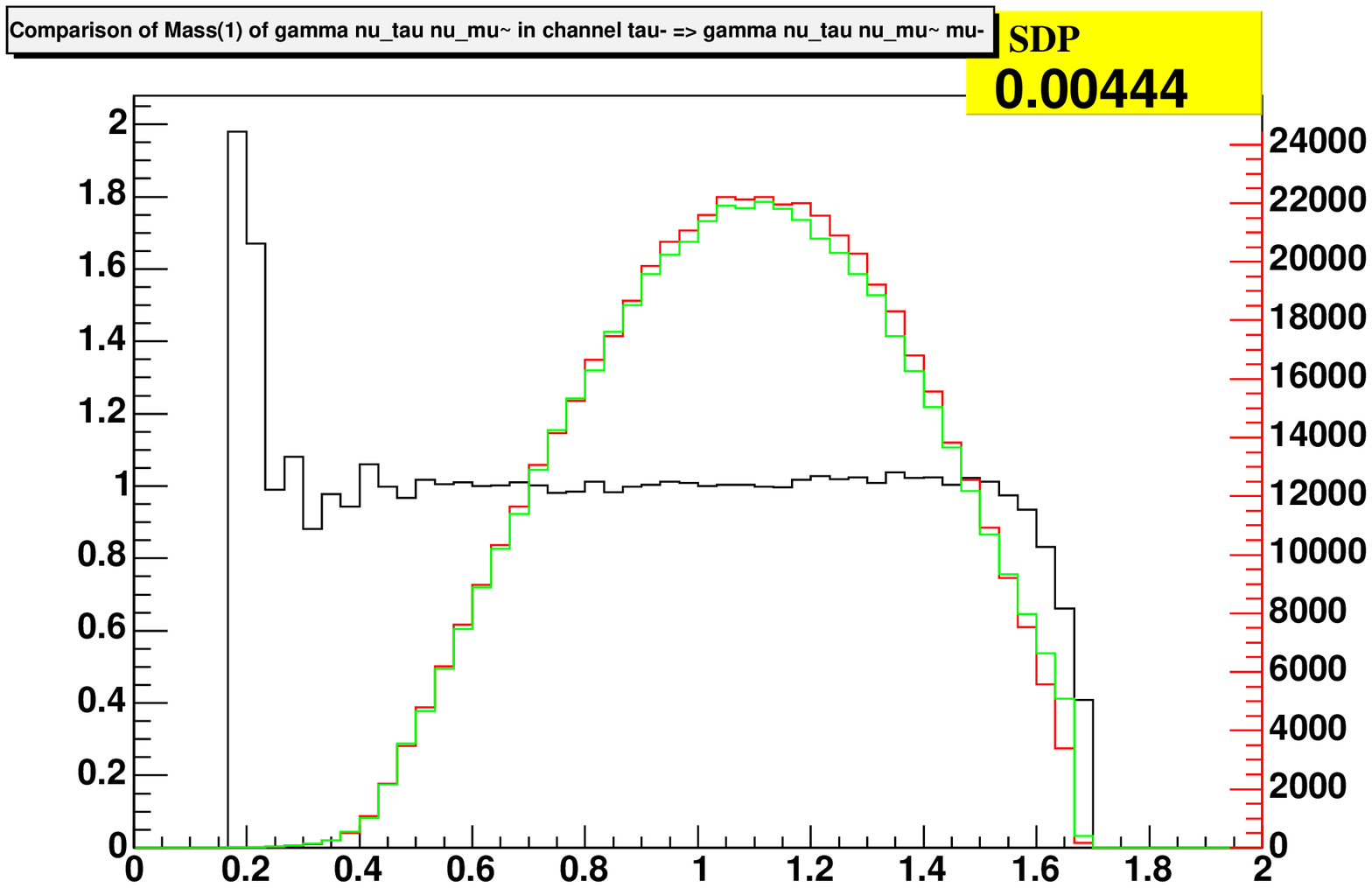}\end{center}\end{minipage}%
\end{center}
\end{figure*}

We have however exploited  the opportunity of having another comparison
with  full $O(\alpha)$ Matrix Element generator, that is TAUOLA \cite{Jezabek:1991qp} for 
leptonic $\tau$ decays.
The results of comparison in $\tau^- \rightarrow \mu^- \bar{\nu}_\mu \nu_{\tau} (\gamma)$
decay channel are presented in Fig. \ref{fig:O1ResultsTAUOLAvsPHOTOSmu}.
The overall difference was quantified as 0.11\%.
The agreement between PHOTOS and TAUOLA results is excellent.
In the $\tau^- \rightarrow e^- \bar{\nu}_e \nu_{\tau}(\gamma)$ channel, 
the agreement is  equally good.

\subsection{\label{sub:Iterating-emission-kernels}Iterating emission kernels}

In this subsection we cover the comparisons of PHOTOS with 
other generators for multiple-photon generation. 
In multiple-photon mode, PHOTOS algorithm iterates the
single-photon emission kernel, which precision was established
in the previous subsection.
The main questions to be answered by the group of tests presented here were
the ones concerning the convergence of the PHOTOS iterative algorithm 
for photon emission to solutions implemented in other generators. 

\begin{figure*}

\caption{\label{fig:O1ResultsKKvsKoralZ} Comparison of 
predictions from KKMC ($O(\alpha^2)$ ME with exponentiation
at the level of spin amplitudes) 
and KORALZ ($O(\alpha^2)$ ME with exponentiation)
 in the $Z^0 \rightarrow \mu^+ \mu^- (\gamma)$ channel, $E_{test}=1.0$ GeV,
{\em test1} used for analysis.
The plot presents the distribution of invariant mass of $\mu^+ \mu^-$ pair 
coming from KKMC (in red, or darker-grey) and KORALZ (in green, or lighter-grey); 
the black line is the ratio of the two normalised distributions. The effects due
to different types of exponentiation are small, albeit noticeable, and
constitute 0.066\% overall difference.}

\begin{center}\begin{minipage}[c]{1.0\textwidth}%
\begin{center}~ \\
 \begin{tabular}{|c|c|c|c|c|}
\hline 
GENERATOR&\multicolumn{2}{c|}{Branching ratio}&\multicolumn{2}{c|}{Max SDP} \tabularnewline
$n$ photons $\to$ &0&1&0&1\tabularnewline
\hline
\hline 
\textbf{KKMC}&\textbf{0.83918}&\textbf{0.16082}&\multicolumn{2}{c|}{}\tabularnewline
\hline 
KORALZ&0.83984&0.16016&0&0.0021\tabularnewline
\hline
\end{tabular} \\
~\\
~\\
\includegraphics[%
  width=0.60\textwidth,
  keepaspectratio]{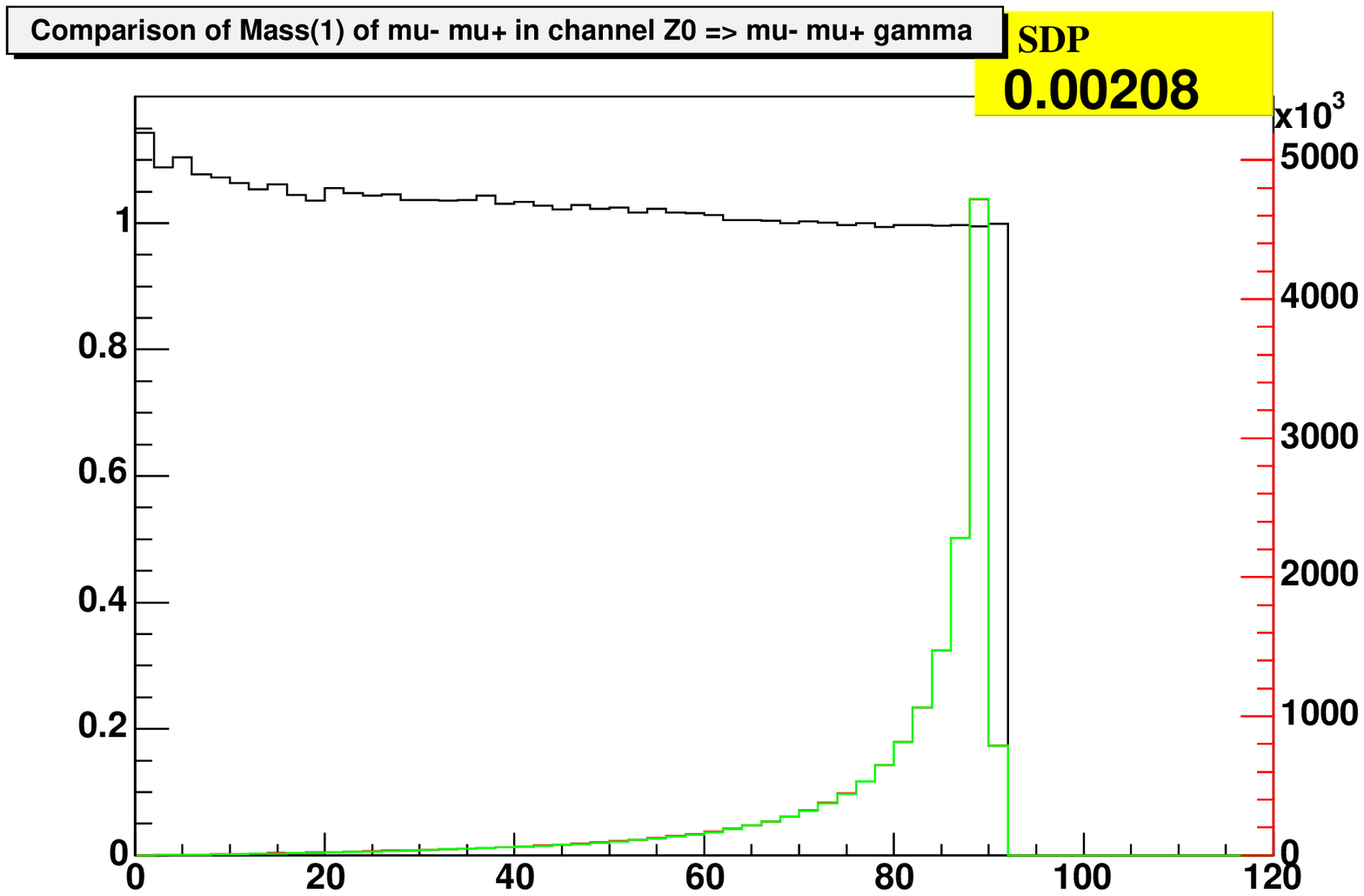}\end{center}\end{minipage}%
\end{center}
\end{figure*}

The KKMC Monte Carlo program \cite{kkcpc:1999} played an important role 
in multiple-photon comparisons: it is the only available Monte Carlo 
that implements a complete $O(\alpha^{2})$ matrix element
for two hard photon emission in the $Z\rightarrow\mu^+\mu^- n(\gamma)$ channel. 
Because of its superior quality with respect to KORALZ  \cite{koralz4:1999}
(which implements $O(\alpha^{2})$ ME with more approximations)
it was used as the reference generator throughout all tests 
in this channel.
Before continuing with tests of PHOTOS let us first compare 
the KORALZ and KKMC Monte Carlo programs to assess the impact of the
presence of second-order terms in the calculations.
The results of comparisons performed using {\it test1} are presented in 
Fig.~\ref{fig:O1ResultsKKvsKoralZ}.
The overall difference is below a per-mil level;
however, with samples of $10^8$ events the contribution from those $O(\alpha^2)$ 
terms that are missing in KORALZ is already visible. As will be shown later, 
the differences are in fact at the level of (and are comparable in shapes to) 
the differences observed between KORALZ and PHOTOS. 

We start the tests of PHOTOS from comparisons for the benchmark  decay
$Z \to \mu^+ \mu^- n(\gamma)$. 
The complete summary of results is presented in Table \ref{fig:Z0Multi-tab}
(as usual, complete results, with MC-TESTER booklets are available
from the web page \cite{tauolaphotos}).
\begin{table}
\caption{\label{fig:Z0Multi-tab} Summary of multiple-photon comparisons
in $Z^{0}\rightarrow\mu^{+}\mu^{-} n(\gamma)$ channel (${Z^0}$ at resonance 
peak, CMS rest-frame);  $E_\mathrm{test}=1.0$ GeV. KKMC was used as a reference generator.
For KKMC--KORALZ comparison the overall difference is $0.00066$,
for KKMC--PHOTOS EXP it is $0.00081$, for KKMC--KKMC $O(1)_\mathrm{EXP}$ it is $0.00137$
(all differences are for {\it test1}) 
}
\begin{center}
\small
\begin{tabular}{|c|c|c|c|c|c|c|c|c|c|c|}
\hline 
GENERATOR & \multicolumn{5}{c|}{Branching ratio} & \multicolumn{5}{c|}{Max SDP}         \tabularnewline
\hline 
 & \multicolumn{2}{c|}{test1} & \multicolumn{3}{c|}{test2} & \multicolumn{2}{c|}{test1} & \multicolumn{3}{c|}{test2} \tabularnewline
$n$ photons $\to$ & 0 & 1 & 0 & 1 & 2 & 0 & 1 & 0 & 1 & 2                         \tabularnewline
\hline
\hline 
\textbf{KKMC }& \textbf{.83918} & \textbf{.16082} & \textbf{.83918} & \textbf{.14816} & \textbf{.01266} & \multicolumn{5}{c|}{} \tabularnewline
\hline 
KORAL Z     & .83984 & .16016 & .83984 & .14771 & .01244 & 0 & .0021 & 0 & .0012 & .0012 \tabularnewline
\hline 
PHOTOS O(2) & .83925 & .16075 & .83925 & .14630 & .01445 & 0 & .0067 & 0 & .0035 & .0122 \tabularnewline
\hline 
PHOTOS O(3) & .83832 & .16168 & .83832 & .14889 & .01280 & 0 & .0038 & 0 & .0025 & .0080 \tabularnewline
\hline 
PHOTOS O(4) & .83836 & .16164 & .83836 & .14871 & .01293 & 0 & .0040 & 0 & .0027 & .0058 \tabularnewline
\hline 
PHOTOS EXP  & .83837 & .16163 & .83837 & .14868 & .01295 & 0 & .0041 & 0 & .0023 & .0092 \tabularnewline
\hline
\hline 
KKMC        &        &        &        &        &       &   &       &   &       &       \tabularnewline
$O(1)_\mathrm{EXP}$&     &        & .83781 & .14881 & .01338 &   &       & 0 & .0099 & .0467 \tabularnewline
\hline
\end{tabular}
\end{center}
\end{table}
We tested PHOTOS running with various options for photon multiplicity: 
fixed (two to four) order and the exponentiated versions
against the reference KKMC generator.

We observed that adding subsequent iterations of photon-emission kernel 
improves the agreement between PHOTOS and KKMC.
The difference in results between quartic-iteration and exponentiated version 
of PHOTOS are negligible; however, the exponentiated version is technically superior,
because it may work with much lower value of the infrared cut-off
parameter (see the appendix for details).

It is striking that the agreement between the exponentiated versions of
PHOTOS and KKMC is best if the full $O(\alpha^{2})$ exponentiated matrix element
is used in KKMC; if KORALZ with exponentiated $O(\alpha)$ is used%
\footnote{These results are not included in this paper.} 
or the matrix element in KKMC is downgraded to exponentiated $O(\alpha)$ 
only (see the last row in the table), the agreement is not as perfect, 
but it is still good enough for any application we can imagine at present. 

The dominant differences are visible in the branching 
fractions for channels with distinct numbers of photons with energies above $E_\mathrm{test}$. 
The pattern of differences may indicate that leading terms summed by
an iterated PHOTOS kernel give better precision than any first-order exponentiated
ME generator for this process. 

\begin{figure*}

\caption{\label{fig:Acoplanarity1}Acoplanarity distributions (in the 
$Z^0 \rightarrow \mu^- \mu^+ \gamma \gamma$ channel) produced
by the KKMC generator running in exponentiated $O(\alpha^2)$ mode
(in red, or darker-grey) and exponentiated $O(\alpha)$  mode
(in green, or lighter grey); the ratio of the two 
distributions plotted in black. For more details see subsection \ref{sub:Iterating-emission-kernels}.}

\begin{center}\includegraphics[%
  width=0.70\textwidth,
  keepaspectratio]{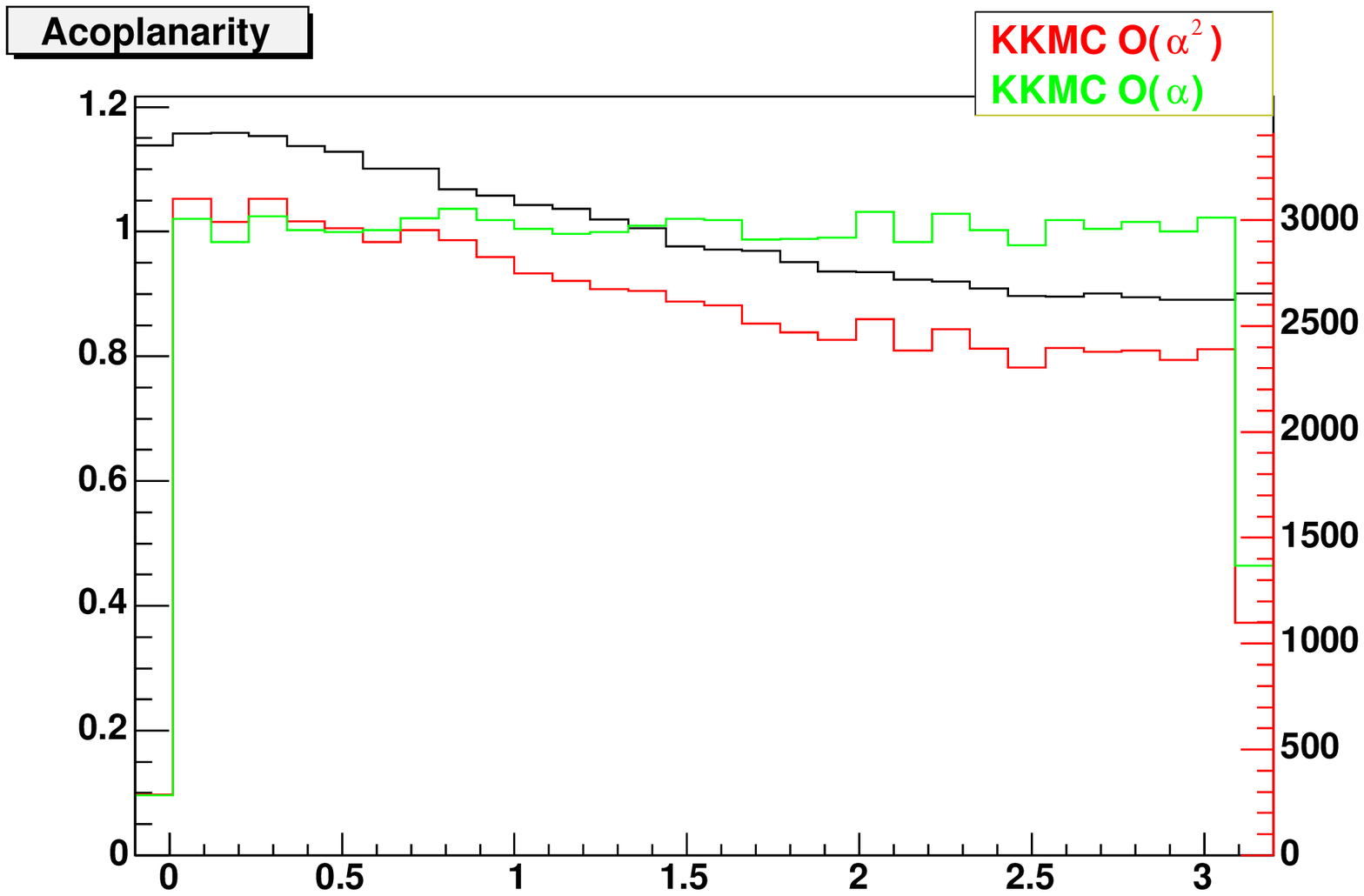}\end{center}
\end{figure*}

To investigate this hypothesis further, we have turned our attention to  
acoplanarity distributions  obtained from  PHOTOS and KKMC.
One of the effects brought by NNLO terms of $O(\alpha^2)$ ME
is the asymmetry in the acoplanarity distributions. 
To define acoplanarity, the
two planes are spanned on momenta vectors of $\mu^-$ and two hardest photons.
In Fig.~\ref{fig:Acoplanarity1} we show
acoplanarity distributions from the KKMC generator with $O(\alpha^2)$
and $O(\alpha)$ exponentiated matrix-element  modes%
\footnote{In order to make the effect visible, we used the following cuts: only
the events with both photons in the same hemisphere as $\mu^-$
and having $p_{t}>1.5$ GeV were recorded in the histogram.}.
As expected, the distribution is flat for the $O(\alpha)$ exponentiated mode and 
asymmetry appears for the $O(\alpha^2)$ exponentiated mode only.

\begin{figure*}

\caption{\label{fig:Acoplanarity2}
Acoplanarity distributions (in the
$Z^0 \rightarrow \mu^- \mu^+ \gamma \gamma$ channel) produced
by the KKMC generator running in exponentiated $O(\alpha^2)$ mode
(in red, or darker-grey) and the exponentiated PHOTOS algorithm
(in green, or lighter grey); the ratio of the two 
distributions is plotted as dotted black curve. 
For more details see subsection \ref{sub:Iterating-emission-kernels}.}

\begin{center}
\includegraphics[%
  width=0.70\textwidth,
  keepaspectratio]{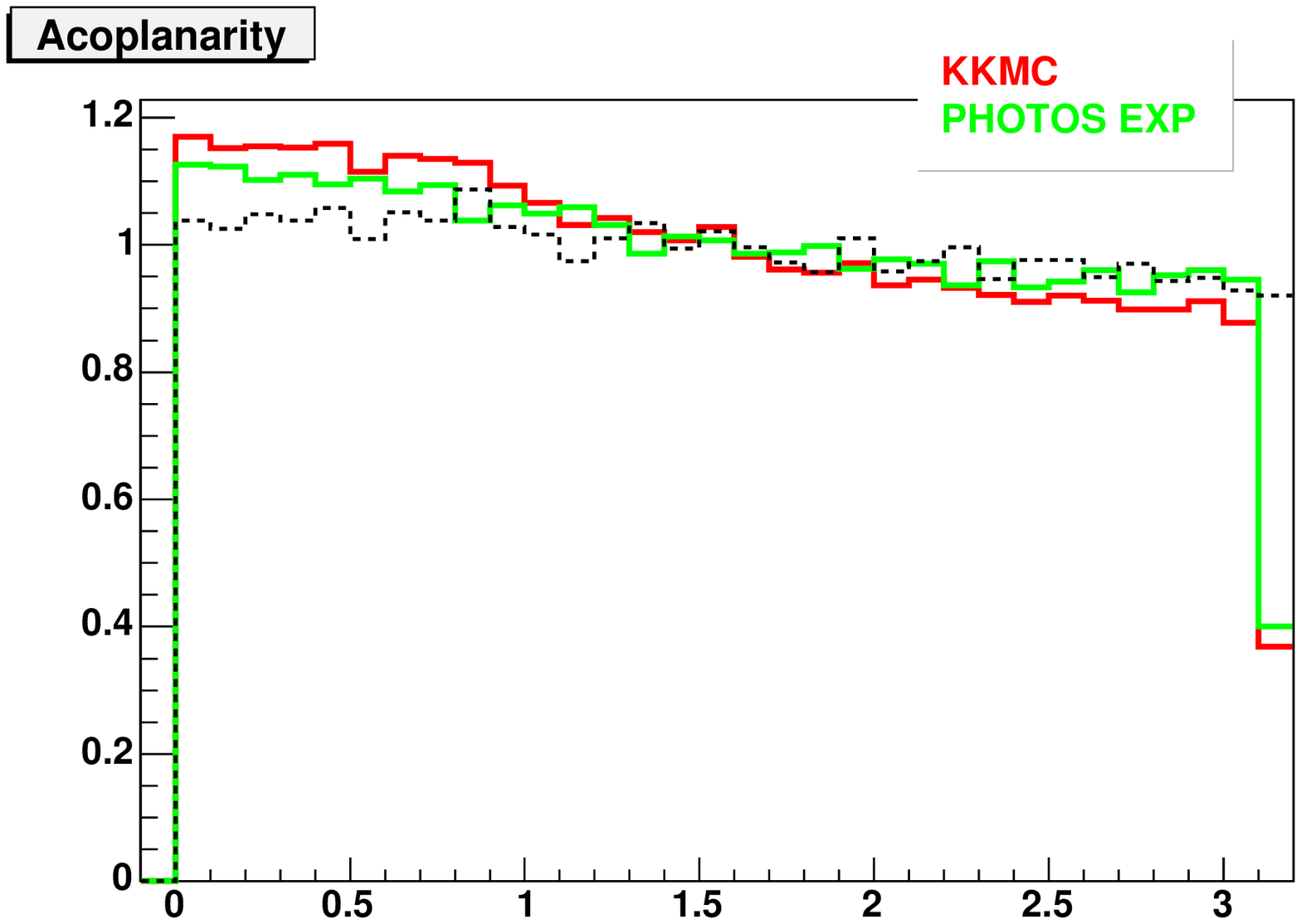}\\
\end{center}
\end{figure*}

In Fig.~\ref{fig:Acoplanarity2} we present the acoplanarity distributions
from the exponentiated $O(\alpha^2)$ ME mode of KKMC and 
exponentiated version of PHOTOS. 
Surprisingly, PHOTOS
seems to reproduce the bulk of this NNLO effect! This rather subtle effect
requires investigation. PHOTOS does not use any kind of phase-space ordering, but,
at the time when the second photon is generated, the momentum of the previous 
one is changed and so the correlations between the directions of two photons appear. 
We shall not elaborate further on this effect here; however,
we believe that further enhancements of precision of the PHOTOS algorithm in this 
aspect may be possible, for instance by introducing an asymmetry 
in the generation of a photon polar angle (uniform distribution 
is currently used in emission kernels).

\begin{figure*}
\caption{\label{fig:ResultsWINHACvsPHOTOS} Multiple-photon comparison of 
WINHAC and PHOTOS (both running in the exponentiated mode)
in the $W^{+}\rightarrow\mu^{+}\nu_{\mu} n(\gamma)$ channel, $E_\mathrm{test}=1.0$ GeV, 
analysed using {\em test1}.
The plot (of largest SDP) presents the distribution of invariant mass of the
$\mu^+ \gamma$ pair coming from WINHAC (in red, or darker-grey)
and PHOTOS (in green, or lighter-grey);
the red and green lines practically overlap.
The black line is the ratio of the two normalized distributions.}
\begin{center}\begin{minipage}[c]{1.0\textwidth}%
\begin{center}\begin{tabular}{|c|c|c|c|c|}
\hline GENERATOR&\multicolumn{2}{c|}{Branching ratio}&\multicolumn{2}{c|}{Max SDP}\tabularnewline
$n$ photons $\to$&0&1&0&1 \tabularnewline
\hline
\hline 
\textbf{WINHAC (EXP)}& \textbf{0.92771}& \textbf{0.07229}& \multicolumn{2}{c|}{}\tabularnewline
\hline 
PHOTOS (EXP)&0.92748&0.07252&0&0.0029\tabularnewline
\hline
\end{tabular}\\
~\\
~ \\
\includegraphics[%
  width=0.60\textwidth]{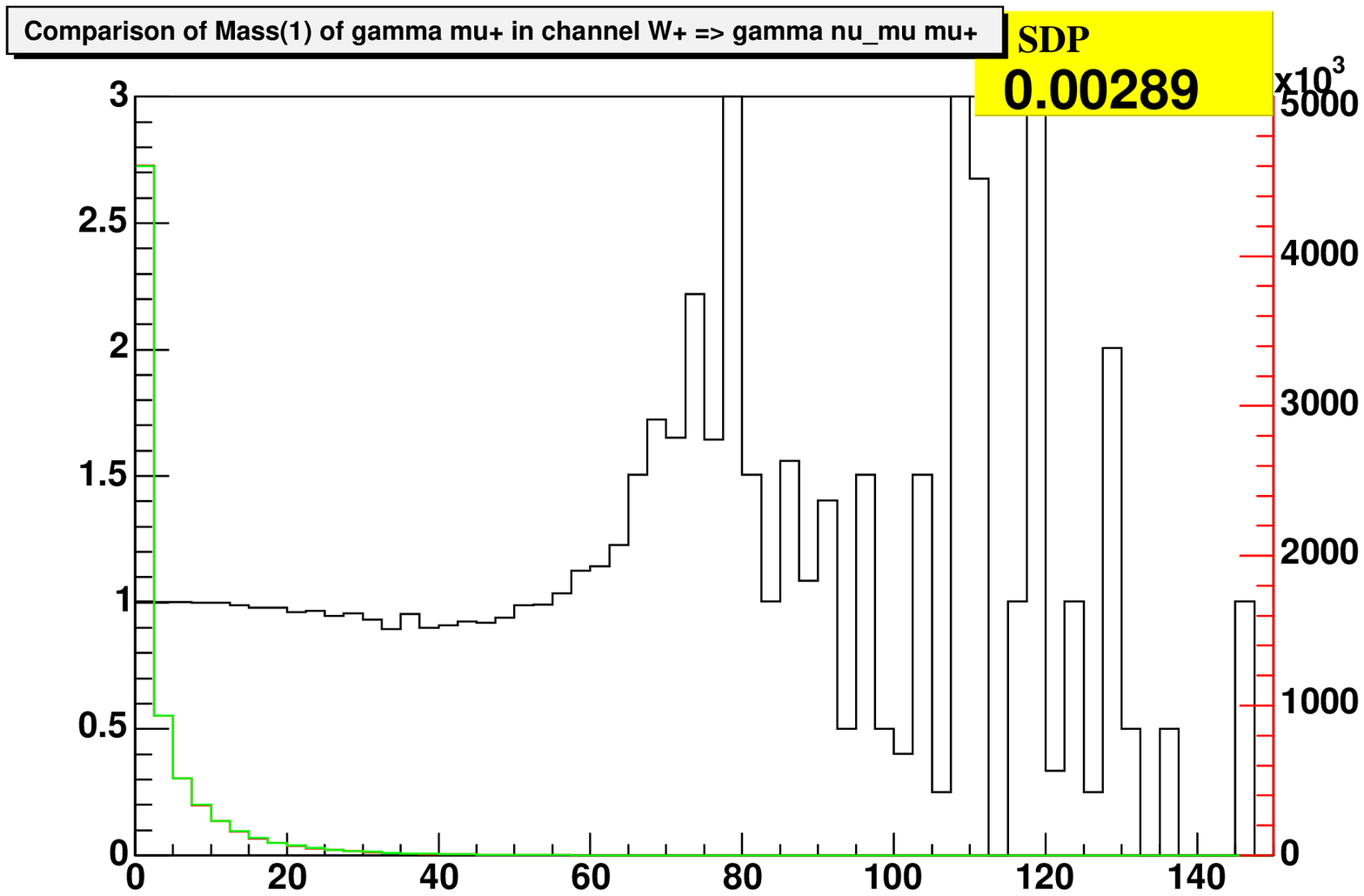}\end{center}\end{minipage}%
\end{center}
\end{figure*}

Let us now present the results of comparisons of exponentiated mode
of PHOTOS and full $O(\alpha)$ exponentiated ME 
predictions of WINHAC for $W \rightarrow l \bar{\nu}_l (\gamma)$ channels.
Both programs were running in exponentiated mode. 
The complete summary of results is presented in Table \ref{fig:WMulti-tab}. 
\begin{table}
\caption{\label{fig:WMulti-tab} Summary of comparisons of the exponentiated
algorithm of PHOTOS and WINHAC $O(\alpha)$ exponentiated 
for leptonic $W$ decays. For more details, see the text.}
\begin{center}
\small
\begin{tabular}{|c|c|c|c|c|c|c|c|c|c|c|}
\hline 
GENERATOR&
\multicolumn{5}{c|}{Branching ratio} & \multicolumn{5}{c|}{Max SDP}\tabularnewline
\hline 
& \multicolumn{2}{c|}{test1} & \multicolumn{3}{c|}{test2} & \multicolumn{2}{c|}{test1} & \multicolumn{3}{c|}{test2} \tabularnewline
$n$ photons $\to$ & 0 & 1 & 0 & 1 & 2 & 0 & 1 & 0 & 1 & 2\tabularnewline
\hline
\hline 
\multicolumn{11}{|c|}{ 
$ W^{+}\rightarrow\mu^{+}\nu_{\mu}(\gamma)$ at $W$ mass, $E_\mathrm{test}=1.0$ GeV}\tabularnewline
\hline
\hline 
\textbf{WINHAC EXP} & \textbf{.92771} & \textbf{.07229} & \textbf{.92771} & \textbf{.07007} & \textbf{.00222} & \multicolumn{5}{c|}{} \tabularnewline
\hline 
PHOTOS EXP & .92748 & .07252 & .92748 & .07016 & .00236 & 0 & .0029 & 0 & .0025 & .0023 \tabularnewline
\hline
\hline 
\multicolumn{11}{|c|}{
$W^{+}\rightarrow\mu^{+}\nu_{\mu}(\gamma)$ at $W$ mass, $E_\mathrm{test}=5.0$ GeV}\tabularnewline
\hline
\hline 
\textbf{WINHAC EXP} & \textbf{.96491} & \textbf{.03509} & \textbf{.96491} & \textbf{.03473} & \textbf{.00036} & \multicolumn{5}{c|}{} \tabularnewline
\hline 
PHOTOS EXP & .96470 & .03530 & .96470 & .04488 & .00042 & 0 & .0060 & 0 & .006 & .047 \tabularnewline
\hline
\hline 
\multicolumn{11}{|c|}{
$W^{+}\rightarrow e^{+}\nu_{e}(\gamma)$ at $W$ mass, $E_\mathrm{test}=1.0$ GeV}\tabularnewline
\hline
\hline 
\textbf{WINHAC EXP} & \textbf{.86196} & \textbf{.13804} & \textbf{.86196} & \textbf{.12943} & \textbf{.00862} & \multicolumn{5}{c|}{} \tabularnewline
\hline 
PHOTOS EXP & .86205 & .13795 & .86205 & .12909 & .00886 & 0 & .0022 & 0 & .0019 & .0094 \tabularnewline
\hline
\hline 
\multicolumn{11}{|c|}{
$W^{+}\rightarrow e^{+}\nu_{e}(\gamma)$ at $W$ mass, $E_\mathrm{test}=5.0$ GeV}\tabularnewline
\hline
\hline 
\textbf{WINHAC EXP} & \textbf{.93083} & \textbf{.06917} & \textbf{.93083} & \textbf{.06763} & \textbf{.00153} & \multicolumn{5}{c|}{} \tabularnewline
\hline 
PHOTOS EXP & .93087 & .06913 & .93087 & .06747 & .00166 & 0 & .0046 & 0 & .0041 & .0255 \tabularnewline
\hline
\end{tabular}
\end{center}
\end{table}
The details of one of these comparisons are presented
in Fig.~\ref{fig:ResultsWINHACvsPHOTOS}.
The overall agreement is better than per mil. 
The maximum value of SDP in configurations with two hard photons 
are larger for some tests (reaching the level of 0.0255 in the last 
row of the table); however, this is a channel with rather small branching 
fraction. 

\begin{figure*}

\caption{\label{fig:O1ResultsTAUOLAvsPHOTOSEXPmu}
Comparison of full $O(\alpha)$ in TAUOLA 
and exponentiated version of PHOTOS for 
$\tau^{-}\rightarrow\mu^{-}\nu_{\tau}\bar{\nu}_\mu(\gamma)$ channel,
 $E_\mathrm{test}=0.05$ GeV, {\em test1} is used.
The plot below presents the distribution
of invariant mass of $\nu_{\tau}\bar{\nu}_\mu\gamma$ from TAUOLA 
(in red, or darker-grey) and PHOTOS (in green, or lighter-grey); the black curve is the ratio of the two
normalized distributions.}
\begin{center}\begin{minipage}[c]{1.0\textwidth}%
\begin{center}\begin{tabular}{|c|c|c|c|c|}
\hline 
GENERATOR&\multicolumn{2}{c|}{Branching ratio}&\multicolumn{2}{c|}{Max SDP}\tabularnewline
$n$ photons $\to$&0&1&0&1\tabularnewline
\hline
\hline 
\textbf{TAUOLA}&\textbf{0.98916}&\textbf{0.01084}&\multicolumn{2}{c|}{}\tabularnewline
\hline 
PHOTOS EXP & 0.98935&0.01065&0&0.0047\tabularnewline
\hline
\end{tabular}\\
~\\
~\end{center}

\begin{center}\includegraphics[%
  width=0.60\textwidth,
  keepaspectratio]{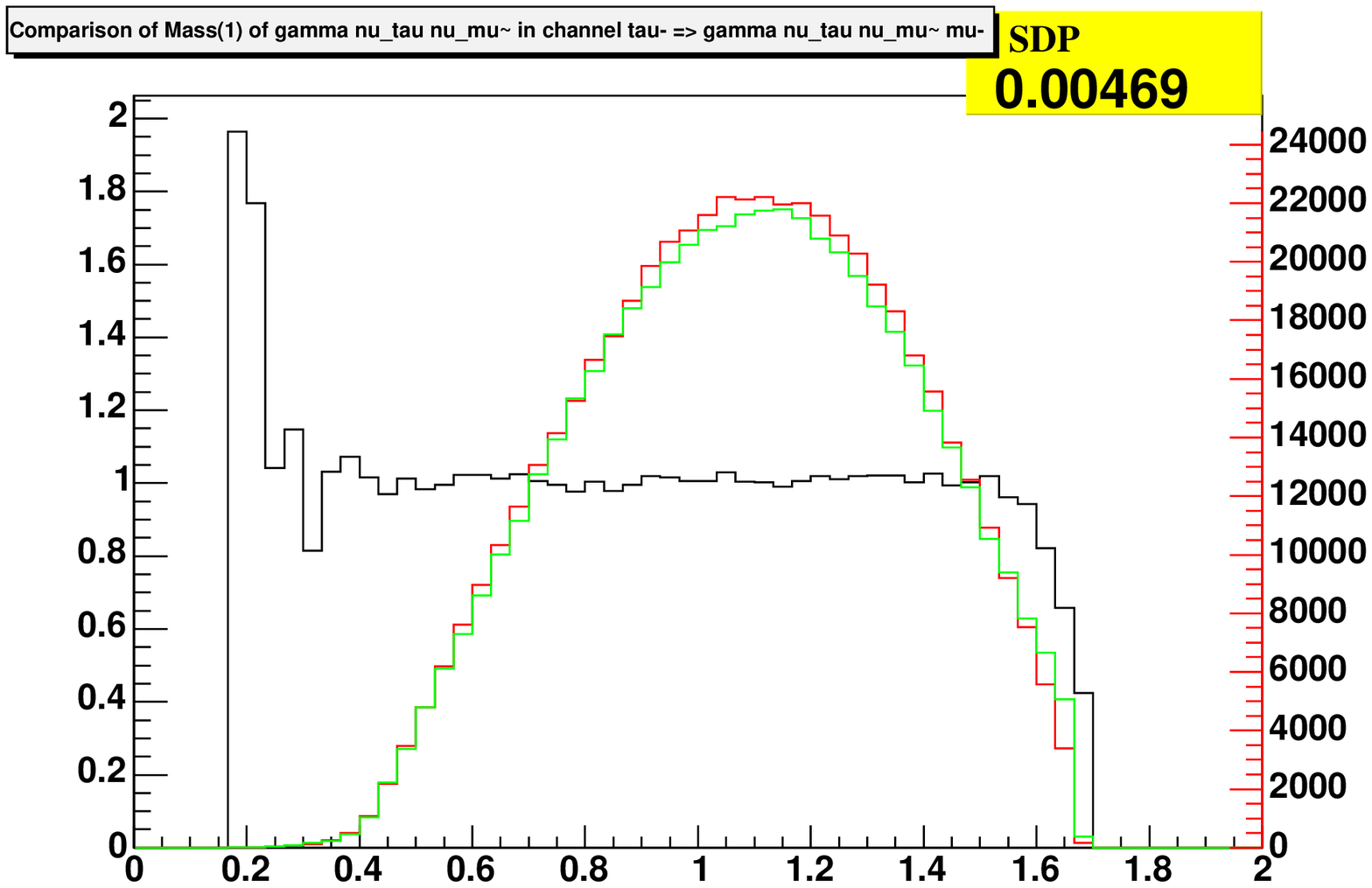}\end{center}\end{minipage}%
\end{center}
\end{figure*}

Finally, let us turn to  $\tau^- \to \mu^- \nu_{\tau} \bar{\nu}_\mu (\gamma)$ 
decay.  The comparison of predictions from the exponentiated version of 
PHOTOS and full $O(\alpha)$ ME of TAUOLA are presented in 
Fig.~\ref{fig:O1ResultsTAUOLAvsPHOTOSEXPmu}. 
The overall agreement is better than 0.02\%.

It is interesting to compare the results presented in 
Fig.~\ref{fig:O1ResultsTAUOLAvsPHOTOSEXPmu} with the results in 
Fig.~\ref{fig:O1ResultsTAUOLAvsPHOTOSmu}, where exact TAUOLA is compared 
with single-photon mode of PHOTOS. One can notice that 
the difference between the single-photon emission mode in PHOTOS and TAUOLA 
is similar to the difference between exponentiated mode of PHOTOS and TAUOLA.
This indicates that the dominant source of differences is due to non-leading $O(\alpha)$ terms
that are missing in PHOTOS (and present in full $O(\alpha)$ ME of TAUOLA).
Predictions of the single-photon mode of TAUOLA are more precise than PHOTOS predictions,
even with exponentiation.
Nevertheless, it must be realized that as TAUOLA does not implement exponentiation,
it is limited to the generation of (at most) single hard
photon emission. 
For cases where the topologies of configurations with two or more photons 
are important, the exponentiated algorithm implemented by PHOTOS will be more appropriate.

\subsection{\label{sub:photos-high-energies}Tests of PHOTOS at very high energies}

Because of numerical instabilities, the use of PHOTOS at LHC energies of 14 TeV
was technically limited before, by a restriction on minimal $\frac{E^\mathrm{min}_{\gamma}}{M}$ ratio
($M$ being the mass of the decaying particle, $E^\mathrm{min}_{\gamma}$ the minimal energy of generated photons).
The phase space for generation had to be significantly reduced and 
photons of even moderate energies could not be generated.

We verified that the exponentiated version of the algorithm, armed with
the new kinematics-correction routine implemented in PHOTOS 2.13,
allows it to be used to generate bremsstrahlung
in decays of very energetic particles at the energy scales
that indeed require very low cut on the minimal photon energy.

We performed studies using KKMC as a ``host'' generator for PHOTOS
to produce leptonic $Z$-decays where the mass of the produced $Z/\gamma$ 
intermediate state was of order 2 TeV.
The value of the infrared cut-off parameter $\frac{E^\mathrm{min}_{\gamma}}{M}$
was lowered down to the value of $\sim 10^{-7}$, which allowed the
generation of relatively soft photons, and high-statistics
runs were completed without encountering numerical stability problems.
Again, we have found excellent agreement between the exponentiated version
of PHOTOS and the exponentiated $O(\alpha^2)$ KKMC.
As usual, complete numerical results can be found on the web page 
\cite{tauolaphotos}.

\subsection{\label{sub:photos-B}Recent developments : semileptonic $K$ decays
and universal interference weight}

Recently our attention was drawn to a use of PHOTOS in an estimation of
the effects of QED radiative corrections in high-precision measurements
of $K$,  $B$ and $D$ meson decays. The systematic error of results given by PHOTOS 
is important for measurements of elements of Cabibbo--Kobayashi--Maskawa matrix
of quark mixing, because  QED corrections affect the acceptance.
The insufficient precision of PHOTOS observed in ref.~\cite{Andre:2004tk}
was of no surprise: for the case of semileptonic $K$ decays, 
PHOTOS was prepared for use in the ``crude'' mode only, 
suitable for full phase space coverage of bremsstrahlung for detector 
studies, but not for shape estimations. The most important missing
terms in calculations performed by PHOTOS in this case were those
from QED interference.
 
Thanks to the technical development performed in PHOTOS, 
and partially described already in this paper, it was rather 
straightforward to implement the universal interference weight for any
decay channel, as expressed by formula (17) of \cite{Barberio:1994qi}:
\begin{eqnarray}
W_{multi}= { \sum_\varepsilon \Bigl|
Q_1{q_1 \cdot \varepsilon \over q_1 \cdot k }
+Q_2{q_2 \cdot \varepsilon \over q_2 \cdot k } + ...
\Bigr|^2
\over
  \sum_\varepsilon
Q_1^2\Bigl|
 {q_1 \cdot \varepsilon \over q_1 \cdot k } \Bigr|^2
+Q_2^2\Bigl|{q_2 \cdot \varepsilon \over q_2 \cdot k }\Bigr|^2 + ... },
\end{eqnarray}
where $Q_1$, $Q_2, ...$ denote the charges of the decay products,
$q_1$, $q_2, ...$ denote the momenta of the decay products,
$k$ denotes the energy of photon, and a summation is performed over 
photon polarization states denoted by $\varepsilon$.

\begin{figure}
\caption{\label{fig:Kmu3_NewInterf} Impact of the universal interference weight
in PHOTOS on predictions for the angular distribution in $K\rightarrow\mu\nu(\gamma)$ events.
 The distribution of the cosine of the
angle between the charged lepton and the photon generated by older version of PHOTOS,
without the interference weight (plotted as black dots), and PHOTOS 2.13 with
universal interference weight (plotted as red line). This figure should
be compared with figure 8b from \cite{Andre:2004tk}. The majority of discrepancies
between PHOTOS and $O(\alpha)$ generator KLOR seems to be removed
by using the universal interference weight.
}
\begin{center}
\includegraphics[bb=0bp 0bp 567bp 567bp, clip, width=0.50\textwidth, keepaspectratio]{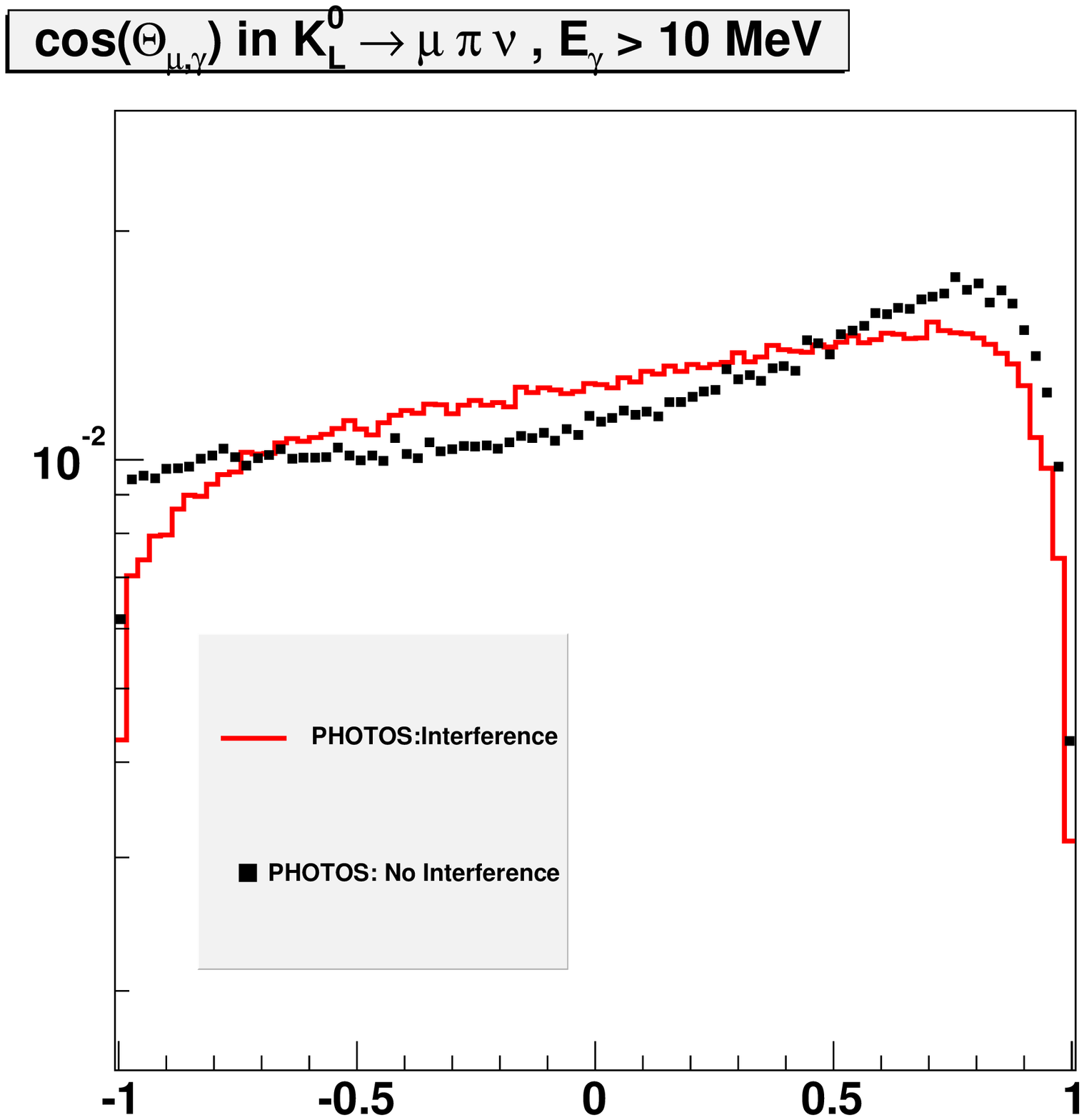}
\end{center}
\end{figure}

Earlier, the interference weight in PHOTOS was calculated from internal
angular variables and not from four-vectors; therefore such universal
implementation was not trivial. 
The impact of the new interference weight on the photon angle distributions in 
$K\rightarrow\mu\nu\pi(\gamma)$ process is presented in Fig.~\ref{fig:Kmu3_NewInterf}.
It seems that the majority of discrepancies between PHOTOS and the first-order
ME generator KLOR have been removed in this way\footnote {For the complete comparison,
see Fig. 8 of \cite{Andre:2004tk} and our comparison web page \cite{tauolaphotos}.}.
The complete study will however be documented elsewhere \cite{LitovMihova}.

\section{\label{sec:Summary-and-Outlook}Summary and Outlook}

We have presented numerical results from a recently upgraded version 
of the PHOTOS Monte Carlo algorithm for radiative corrections
in decays of resonances. 
In the case of $W$ and $Z$ decays, we found the precision of the simulations 
in complete multiple-photon final states at the level of (or better than) $0.1$\%. 
This conclusion originates from numerous comparisons with precision Monte Carlo 
programs: KORALZ, KKMC and WINHAC for simulation of radiative corrections in $Z$ 
and $W$ decays. 
The reliability of those programs -- from theoretical and technical sides -- was well 
established by numerous applications at LEP time, documented in a multitude 
of publications on comparisons with the experimental data (in the case of $Z$
 decay), as well as other programs and calculations.

The results we presented prove numerically that PHOTOS may be used with confidence
to calculate bremsstrahlung in $Z$ and $W$ leptonic decays with very high precision.
In the case of Higgs boson decays and leptonic decays of $\tau$'s,
this precision tag is valid as well, even if the proof is not
as solid, since it relies on the test at first order only.

This conclusion cannot be extended to other
decay channels, without further work involving a study of process-dependent
matrix elements. Nonetheless, precision in the general case is also better now.

Finally, let us stress that the mathematical side of the multiple-photon
algorithm presented here is missing a rigorous proof. Present-day applications
do not justify such an effort. Wherever it was possible, we have verified 
PHOTOS correctness and found that the performance was excellent. In other cases
we were limited by availability of exact matrix elements for single-photon
emission. 

A more thorough explanation of the mathematical side of the algorithm
may, however, be interesting in the future, especially having in mind
applications in QCD. Note that algorithms of PHOTOS do not rely on the conformal 
symmetry of soft photon factors and the phase space of masless particles, 
which is  crucial in the construction of exclusive-exponentiation 
algorithms as used by KKMC, WINHAC or KORALZ. 

From its origin, the algorithm of PHOTOS was developed as FORTRAN77 code, 
even though a C++ implementation was already completed \cite{MsCGolonka}
several years ago. 
Since there was no standard for C++ event record at that time,
it was not published.
Recently, the experimental communities expressed interest in using PHOTOS with 
the HepMC event record \cite{Dobbs:2001ck}. 
This certainly is a valid option for future development.

\section*{Acknowledgements}

We would like to acknowledge inspiring discussions with the authors
of the generators we used for comparisons, and with the users of PHOTOS;
in particular we would like to thank E.~Barberio, M.~Boonekamp, 
S.~Jadach, B.~Kersevan, R.~Kessler, L.~Litov,  W.~Placzek, 
E.~Richter-Was, B.F.L.~Ward. 

The inspiring atmosphere of HERA--LHC and MC4LHC workshops at CERN and
at DESY is also acknowledged.

\newpage


\begingroup\begin{thebibliography}{10}

\bibitem{Barberio:1990ms}
E.~Barberio, B.~van Eijk, and Z.~Was, {\em Comput. Phys. Commun.} {\bf 66}
  (1991)
115.

\bibitem{Barberio:1994qi}
E.~Barberio and Z.~Was, {\em Comput. Phys. Commun.} {\bf 79} (1994)
291--308.

\bibitem{Dobbs:2004qw}
M.~A. Dobbs {\em et al.},
\href{http://www.arXiv.org/abs/hep-ph/0403045}{{\tt hep-ph/0403045}}.

\bibitem{Abazov:2003sv}
{CDF} Collaboration, V.~M. Abazov {\em et al.}, {\em Phys. Rev.} {\bf D70}
  (2004) 092008,
\href{http://www.arXiv.org/abs/hep-ex/0311039}{{\tt hep-ex/0311039}}.

\bibitem{Abbiendi:2003jh}
{OPAL} Collaboration, G.~Abbiendi {\em et al.}, {\em Phys. Lett.} {\bf B580}
  (2004) 17--36,
\href{http://www.arXiv.org/abs/hep-ex/0309013}{{\tt hep-ex/0309013}}.

\bibitem{Abdallah:2003xn}
{DELPHI} Collaboration, J.~Abdallah {\em et al.}, {\em Eur. Phys. J.} {\bf C31}
  (2003) 139--147,
\href{http://www.arXiv.org/abs/hep-ex/0311004}{{\tt hep-ex/0311004}}.

\bibitem{Lai:2004bt}
{NA48} Collaboration, A.~Lai {\em et al.}, {\em Phys. Lett.} {\bf B602} (2004)
  41--51,
\href{http://www.arXiv.org/abs/hep-ex/0410059}{{\tt hep-ex/0410059}}.

\bibitem{Alexopoulos:2004up}
{KTeV} Collaboration, T.~Alexopoulos {\em et al.}, {\em Phys. Rev.} {\bf D71}
  (2005) 012001,
\href{http://www.arXiv.org/abs/hep-ex/0410070}{{\tt hep-ex/0410070}}.

\bibitem{Limosani:2005pi}
{Belle} Collaboration, A.~Limosani {\em et al.},
\href{http://www.arXiv.org/abs/hep-ex/0504046}{{\tt hep-ex/0504046}}.

\bibitem{Aubert:2004te}
{BABAR} Collaboration, B.~Aubert {\em et al.}, {\em Phys. Rev.} {\bf D69}
  (2004) 111103,
\href{http://www.arXiv.org/abs/hep-ex/0403031}{{\tt hep-ex/0403031}}.

\bibitem{Link:2004vk}
{FOCUS} Collaboration, J.~M. Link {\em et al.},
\href{http://www.arXiv.org/abs/hep-ex/0412034}{{\tt hep-ex/0412034}}.

\bibitem{PGZWinprep}
P.~Golonka and Z.~Was, In preparation.

\bibitem{PhDGolonka}
P.~Golonka, {\em In preparation}.
\newblock PhD thesis, Institute of Nuclear Physics, P.A.S., Krakow, 2005?
\newblock Written under supervision of Z. Was, draft will be available at {\tt
  http://cern.ch/Piotr.Golonka/MC/PhD}.

\bibitem{Was:2004dg}
Z.~Was and P.~Golonka,
\href{http://www.arXiv.org/abs/hep-ph/0411377}{{\tt hep-ph/0411377}}.

\bibitem{Was:1994kg}
Z.~Was, Written on the basis of lectures given at the 1993 European School of
  High Energy Physics, Zakopane, Poland, 12-25 Sep 1993,
  \href{http://www.arXiv.org/abs/CERN-TH-7154-94}{{\tt CERN-TH-7154-94}}.

\bibitem{Richter-Was:1994ep}
E.~Richter-Was, {\em Z. Phys.} {\bf C64} (1994)
227--240.

\bibitem{Richter-Was:1993ta}
E.~Richter-Was, {\em Z. Phys.} {\bf C61} (1994)
323--340.

\bibitem{LaJolla2005}
``Radiative corrections in B, D and K meson decays'' workshop and CKM05
  Conference, La Jolla, March 2005,.

\bibitem{kkcpc:1999}
S.~Jadach, Z.~W\c{a}s, and B.~F.~L. Ward, {\em Comput. Phys. Commun.} {\bf 130}
  (2000) 260, Up to date source available from http://home.cern.ch/jadach/.

\bibitem{koralz4:1994}
S.~Jadach, B.~F.~L. Ward, and Z.~W\c{a}s, {\em Comput. Phys. Commun.} {\bf 79}
  (1994) 503.

\bibitem{Eberhard:1989ve}
P.~H. Eberhard {\em et al.}, To appear in the Proceedings of the Workshop on Z
  Physics at LEP, edited by G. Altarelli, R. Kleiss and V. Verzegnassi
  (CERN-89-08 v.1-3) held in Geneva, Switzerland, Feb 20-21 and May 8-9, 1989.
  Published in LEP Physics Wrkshp.1989:v.1:235-266.

\bibitem{Berends:1982ie}
F.~A. Berends, R.~Kleiss, and S.~Jadach, {\em Nucl. Phys.} {\bf B202} (1982)
63.

\bibitem{Berends:1983mi}
F.~A. Berends, R.~Kleiss, and S.~Jadach, {\em Comput. Phys. Commun.} {\bf 29}
  (1983)
185--200.

\bibitem{Jezabek:1991qp}
M.~Jezabek, Z.~Was, S.~Jadach, and J.~H. Kuhn, {\em Comput. Phys. Commun.} {\bf
  70} (1992)
69--76.

\bibitem{Was:2004ig}
Z.~Was,
\href{http://www.arXiv.org/abs/hep-ph/0406045}{{\tt hep-ph/0406045}}.

\bibitem{yfs:1961}
D.~R. Yennie, S.~Frautschi, and H.~Suura, {\em Ann. Phys. (NY)} {\bf 13} (1961)
  379.

\bibitem{Jadach:1991ws}
S.~Jadach, B.~F.~L. Ward, and Z.~Was, {\em Comput. Phys. Commun.} {\bf 66}
  (1991)
276--292.

\bibitem{Hernandez:1990yc}
J.~J. Hernandez {\em et al.}, {\em Phys. Lett.} {\bf B239} (1990)
1--515.

\bibitem{Golonka:2000iu}
P.~Golonka, E.~Richter-Was, and Z.~Was,
\href{http://www.arXiv.org/abs/hep-ph/0009302}{{\tt hep-ph/0009302}}.

\bibitem{Nanava:2003cg}
G.~Nanava and Z.~Was, {\em Acta Phys. Polon.} {\bf B34} (2003) 4561--4570,
\href{http://www.arXiv.org/abs/hep-ph/0303260}{{\tt hep-ph/0303260}}.

\bibitem{Golonka:2003xt}
P.~Golonka {\em et al.},
\href{http://www.arXiv.org/abs/hep-ph/0312240}{{\tt hep-ph/0312240}}.

\bibitem{koralz4:1999}
S.~Jadach, B.~F.~L. Ward, and Z.~Was,
hep-ph/9905205, Computer. Phys. Commun. in print.

\bibitem{Placzek:2003zg}
W.~Placzek and S.~Jadach, {\em Eur. Phys. J.} {\bf C29} (2003) 325--339,
\href{http://www.arXiv.org/abs/hep-ph/0302065}{{\tt hep-ph/0302065}}.

\bibitem{Andonov:2002mx}
A.~Andonov, S.~Jadach, G.~Nanava, and Z.~Was, {\em Acta Phys. Polon.} {\bf B34}
  (2003) 2665--2672,
\href{http://www.arXiv.org/abs/hep-ph/0212209}{{\tt hep-ph/0212209}}.

\bibitem{Davier:2002mn}
{ALEPH} Collaboration, M.~Davier and C.-z. Yuan, {\em eConf} {\bf C0209101}
  (2002) TU06,
\href{http://www.arXiv.org/abs/hep-ex/0211057}{{\tt hep-ex/0211057}}.

\bibitem{Anderson:1999ui}
{CLEO} Collaboration, S.~Anderson {\em et al.}, {\em Phys. Rev.} {\bf D61}
  (2000) 112002,
\href{http://www.arXiv.org/abs/hep-ex/9910046}{{\tt hep-ex/9910046}}.

\bibitem{Golonka:2002rz}
P.~Golonka, T.~Pierzchala, and Z.~Was, {\em Comput. Phys. Commun.} {\bf 157}
  (2004) 39--62,
\href{http://www.arXiv.org/abs/hep-ph/0210252}{{\tt hep-ph/0210252}}.

\bibitem{tauolaphotos}
P.~Golonka and Z.~Was, see \\ http://cern.ch/Piotr.Golonka/MC/PHOTOS-MCTESTER.

\bibitem{Andonov:2002jg}
A.~Andonov {\em et al.},
\href{http://www.arXiv.org/abs/hep-ph/0209297}{{\tt hep-ph/0209297}}.

\bibitem{Andre:2004tk}
T.~C. Andre,
\href{http://www.arXiv.org/abs/hep-ph/0406006}{{\tt hep-ph/0406006}}.

\bibitem{LitovMihova}
P.~Golonka, L.~Litov, E.~Mihova, and Z.~Was, In preparation.

\bibitem{MsCGolonka}
P.~Golonka, ``Photos+ - a C++ implementation of a universal monte carlo
  algorithm for qed radiative corrections in particle's decays'', Master's
  thesis, Faculty of Nuclear Physics and Techniques, AGH University of Science
  and Technology, June, 1999.
\newblock Written under the supervision of Z. Was, available at {\tt
  http://cern.ch/Piotr.Golonka/MC/photos}.

\bibitem{Dobbs:2001ck}
M.~Dobbs and J.~B. Hansen, {\em Comput. Phys. Commun.} {\bf 134} (2001)
41--46.

\end{thebibliography}\endgroup

\providecommand{\href}[2]{#2}

\newpage

\appendix

\section{\label{sec:new-photos-features}New features of PHOTOS version 2.13}

In this appendix let us briefly summarize options and switches available
in version 2.13 of PHOTOS, so it can serve as a reference guide
for librarians and end-users. The complete, technical description 
of the algorithm, with its recent improvements, will be presented 
elsewhere \cite{PGZWinprep,PhDGolonka}.

By default, the new options are not active and  
PHOTOS 2.13 behaves much like PHOTOS 2.0 from 1994 \cite{Barberio:1994qi}. 
It also does not require any changes in old interfaces. 
In most applications it  will produce  results that differ little
with respect to previous versions, with the exception of the case
of interference effects which are now simulated.
As in previous versions, the initialization parameters,
which are subject to user modifications, should be configured in
the subroutine {\tt PHOCIN}.

\subsection{Standard settings}
In Table~\ref{tab:options-std} we summarize all switches used in standard
PHOTOS setups. The names of the options that are new (or need special
care) in version 2.13 are typed in bold font.

\begin{table}
\caption{\label{tab:options-std} Standard PHOTOS switches and configuration parameters. (exp) denotes
the switches and values that are specific to exponentiated mode.}
\begin{center}
\small
\begin{tabular}{|c|c|c|c|}
\hline 
Subroutine &  Parameter       & Default & Description                                       \tabularnewline
\hline 
\hline
\hline
\multicolumn{4}{c}{Options for photon multiplicity selection}                              \tabularnewline
\hline
PHOCIN     & {\tt ISEC }      & .TRUE.  & Switch to enable double emission                  \tabularnewline
\hline
PHOCIN     & {\bf ITRE}   & .FALSE. & Switch to enable triple/quartic emission                  \tabularnewline
\hline
PHOCIN     & {\bf IEXP}   & .FALSE. & Switch to enable exponentiated, multiphoton mode \tabularnewline
\hline 
\hline
\multicolumn{4}{c}{Options related to interference}                                        \tabularnewline
\hline 
PHOCIN     & {\bf INTERF} & .TRUE. & Switch to enable universal interference weight    \tabularnewline
\hline
PHOCIN     & {\tt IFW }       & .TRUE. & Switch to enable dedicated interference weight \tabularnewline
           &                  &        & for $W \to l \nu (\gamma)$ decays              \tabularnewline
\hline
PHOCIN     & {\bf FINT }  & 2.D0   & Maximum interference weight.\tabularnewline
           &                  &        & The value may need to be adjusted to $2^{n-1}$ , where\tabularnewline
	   &                  &        & $n$ is the maximal multiplicity of charged decay \tabularnewline
	   &                  &        & products of any elementary branching \tabularnewline
\hline
\multicolumn{4}{c}{Other technical options}                                                \tabularnewline
\hline 
\hline
PHOCIN     & {\tt CALL PHCORK(n) } &                & Switch to select kinematics correction mode:\tabularnewline
           &                       & n=1            & n=1: no correction                    \tabularnewline
	   &                       &                & n=2: corrects energy from mass                \tabularnewline
	   &                       &                & n=3: corrects mass from energy                \tabularnewline
	   &                       &                & n=4: corrects energy from mass for particles  \tabularnewline
	   &                       &                & up to 0.4 GeV mass, for heavier ones corrects mass \tabularnewline
	   &                       &  n=5 (exp)     & n=5: most complete correction for mother        \tabularnewline
	   &                       &                & and daughter particles; required for {\it (exp)}\tabularnewline
\hline
PHOCIN     & {\tt XPHCUT}          & 0.01           & Infrared cut-off parameter: minimal photon energy   \tabularnewline
           &                       & 1D-7 (exp) & expressed as fraction of decaying particle mass  \tabularnewline
\hline
PHOCIN     & {\tt IFTOP}           & .TRUE.         & Switch to enable emission in the \tabularnewline
           &                       &                & hard process $gg (q\bar{q}) \to t \bar{t} $\tabularnewline
\hline 
\end{tabular}
\end{center}
\end{table}

\subsubsection*{Exponentiated multiple-photon mode}
The exponentiated mode has been implemented in version 2.09
of PHOTOS; it may be activated using the {\tt IEXP} flag.
By eliminating the problem of negative probabilities we were 
able to lower the default value for {\tt XPHCUT} parameter to 
$10^{-7}$ for that mode. 
Nevertheless, multiple iteration of emission kernel puts 
higher demands on numerical precision: aggregating rounding errors 
 and initially insignificant numerical deficiencies
in energy-momentum conservation of the input data, 
may lead to significant problems with numerical stability.
Therefore a new option for kinematics corrections of the data in the event 
record was implemented\footnote{ This is the extension
of the kinematic correction schemes described in subsection 7.2.1 of \cite{Golonka:2003xt}.
}. The new option, activated with {\tt CALL PHCORK(5)} is active by default for the exponentiated 
mode. It corrects  momenta four-vectors of all particles taking part in photon generation:
at first, the energies of all daughter particles are corrected
in such a way that $E_i^2=p_i^2 + M_i^2$;
then the four-momentum of the mother-particle is recalculated to ensure energy-momentum 
conservation.

\subsubsection*{Universal interference weight}
Before version 2.13 the  QED interference was implemented in PHOTOS
for dedicated decay channels only, namely neutral particle decaying into
a pair of  opposite-charge particles
\footnote {This requirement was relaxed  slightly in PHOTOS 2.07.
Since then the pair of decay products could be of different masses.
See subsection 7.2.2 of \cite{Golonka:2003xt}  for details.
}. 
For other channels\footnote {Except for leptonic $W$ decays, where a dedicated 
interference weight was implemented in version 2.07 (see subsection 7.2.4 of 
\cite{Golonka:2003xt}).} the interference effects were not taken into account so far.

Starting from version 2.13 a new, universal algorithm calculating interference effects
for all processes is present in PHOTOS. By default, as in previous versions, this
algorithm (implemented internally as the Monte Carlo interference weight) is active and
may be controlled using {\tt INTERF}  flag. 
To use this new interference weight, it is required that the value of 
the {\tt FINT} parameter be adjusted in such a way that the maximum  weight 
in the Monte Carlo rejection is increased to $2^{n-1}$, where $n$ denotes 
the maximal expected multiplicity of charged particles in decays
processed by PHOTOS.

An omission in adjusting this parameter often results in PHOTOS stopping with an error message
indicating too high a value of summary weight. On the other hand, putting the
value of {\tt FINT} too high may significantly increase the CPU time consumed by
PHOTOS because of Monte Carlo rejection not being effective.
An option for having  the actual value of {\tt FINT} automatically calculated for 
each event is being considered for implementation in the future.

\subsection{Advanced settings}
In Table \ref{tab:options-adv} we summarize the switches and options dedicated for advanced users.
We strongly recommend to contact the authors before these settings are altered.

\begin{table}
\caption{\label{tab:options-adv} Advanced PHOTOS switches and configuration parameters}
\begin{center}
\small
\begin{tabular}{|c|c|c|c|}
\hline 
Subroutine &  Parameter       & Default & Description                                       \tabularnewline
\hline 
\hline
\hline
PHTYPE     & {\bf IFOUR}  & .TRUE.AND.(ITRE) & Switch to enable quartic emission                 \tabularnewline  
\hline
PHOCIN     & {\bf EXPEPS} & 1D-4 & Residual probability for multiple iteration, \tabularnewline
           &              &      & ``stop criteria'' for crude generation \tabularnewline
\hline
\hline
\multicolumn{4}{c}{Selection of weighting algorithm variant}           \tabularnewline
\hline
PHOCIN     &              &      & Select the variant of weighting algorithm \tabularnewline
PHOMAK     &              &      & code. May be altered by proper commenting \tabularnewline
PHOPRE     &              &      & of code blocks denoted with {\tt VARIANT A}\tabularnewline
PHOENE     &              &      & and {\tt VARIANT B} comments.\tabularnewline
PHOINT     &              &      & For use by experts only.\tabularnewline

\hline 
\end{tabular}
\end{center}
\end{table}

At an intermediate step of development, a more effective Monte Carlo
weight for crude probability in interference calculation
was implemented in PHOTOS 2.07:  instead of doubling the crude 
distribution for photon emission from each charge, a flat, parallel 
channel was added. Although the results were satisfactory, 
this solution turned out to suffer from problems with
stability when used with universal interference weight for more that
three-body decays. 
We therefore reverted the code to the original
implementation. For the purpose of preserving this interesting technical
development and maintaining the compatibility with PHOTOS versions
2.07--2.12, released in 2003--2005, we leave this optimized code 
as an option to be activated by advanced users only. 
The two options are marked in the
code by {\tt VARIANT A} and {\tt VARIANT B} comments, the former being
currently active, the latter being commented-out. Due to elimination
of numerical instabilities, we were able to lower the value of
{\tt XPHCUT} for variant B to $10^{-4}$ and perform important tests.
Such tests were not possible in the default variant A, because
the value of {\tt XPHCUT} has to be kept at $10^{-2}$.

\end{document}